\documentclass[useAMS,usenatbib,usegraphicx]{mn2e}

\usepackage{aas_macros}

\title[The VPOS: a vast polar structure around the MW]{The VPOS: a vast polar structure of satellite galaxies, globular clusters and streams around the Milky Way}
\author[Pawlowski, Pflamm-Altenburg \& Kroupa]{M. S. Pawlowski$^{1}$\thanks{E-mail:
mpawlow@astro.uni-bonn.de}, J. Pflamm-Altenburg$^{1}$ and P. Kroupa$^{1}$\\
$^{1}$Argelander Institute for Astronomy, University of Bonn, Auf dem H\"{u}gel 71, D-53121 Bonn, Germany
}
\begin{document}

\date{Accepted 2012 March 15.  Received 2012 March 6; in original form 2011 August 24}

\pagerange{\pageref{firstpage}--\pageref{lastpage}} \pubyear{2012}

\maketitle

\label{firstpage}

\begin{abstract}
It has been known for a long time that the satellite galaxies of the Milky Way (MW) show a significant amount of phase-space correlation, they are distributed in a highly inclined Disc of Satellites (DoS). We have extended the previous studies on the DoS by analysing for the first time the orientations of streams of stars and gas, and the distributions of globular clusters within the halo of the MW. It is shown that the spatial distribution of MW globular clusters classified as young halo clusters (YH GC) is very similar to the DoS, while 7 of the 14 analysed streams align with the DoS. The probability to find the observed clustering of streams is only 0.3 per cent when assuming isotropy. The MW thus is surrounded by a vast polar structure (VPOS) of subsystems (satellite galaxies, globular clusters and streams), spreading from Galactocentric distances as small as 10 kpc out to 250 kpc.
These findings demonstrate that a near-isotropic infall of cosmological sub-structure components onto the MW is essentially ruled out because a large number of infalling objects would have had to be highly correlated, to a degree not natural for dark matter sub-structures. The majority of satellites, streams and YH GCs had to be formed as a correlated population. This is possible in tidal tails consisting of material expelled from interacting galaxies. We discuss the tidal scenario for the formation of the VPOS, including successes and possible challenges.
The potential consequences of the MW satellites being tidal dwarf galaxies are severe. If all the satellite galaxies and YH GCs have been formed in an encounter between the young MW and another gas-rich galaxy about 10-11 Gyr ago, then the MW does not have any luminous dark-matter substructures and the missing satellites problem becomes a catastrophic failure of the standard cosmological model.
\end{abstract}


\section{Introduction}
\label{sect:introduction}

That the satellite galaxies of the Milky Way fall on a great circle has already been put forward by \citet{LyndenBell1976} before the advent of the standard model of cosmology. In recent years, this substructure defined by the 11 'classical' satellite galaxies\footnote{Carina, Draco, Fornax, Leo I, Leo II, Large and Small Magellanic Clouds, Sagittarius, Sculptor, Sextans and Ursa Minor.} of the MW has been confirmed and named the Disc of Satellites (DoS) \citep*{Kroupa2005,Metz2007}.  Furthermore, the 13 fainter satellite galaxies, mostly detected via the Sloan Digital Sky Survey (SDSS) \citep{York00}, also independently follow this DoS \citep{Metz2009, Kroupa2010}. The satellites of the Andromeda galaxy also show an anisotropic distribution \citep{Koch2006, Metz2007, Metz2009}, see in particular the chain-like distribution of satellites in figure 1 of \citet{Tollerud2011}.

For those satellite galaxies that are fast and close enough such that proper motions could be determined, the orbital poles derived from their spatial kinematics support this anisotropy \citep*{Metz2008}. These suggest that the DoS is rotationally supported as the satellites orbit within this disc.

The satellite galaxies are popularly understood to be luminous dark matter halos which form abundantly in standard cold or warm dark matter (CDM or WDM, respectively) cosmology. However, already independently of the spatial distribution of the satellite galaxies there are problems. For example, \citet{Bovill2011} point out the significant over-abundance of predicted bright satellites compared to observations. As shown by \citet{Strigari2011}, the bright satellite population of the MW is no statistically significant outlier compared to similar host galaxies: the lack of bright satellites is common. If CDM or WDM cosmology were right, how can massive ($\approx 10^{10} M_{\sun}$) dark matter halos remain dark?

The anisotropic distribution of MW satellites has been identified to be a challenge for standard cosmological theory by \citet{Kroupa2005}. Subsequently, confidence in the CDM scenario has been sought in polar structures forming in cosmological simulations. This has for example been shown by \citet{Brook2008}, where torques of a neighbouring galaxy change the orientation of a galactic disc. However, the subsequent formation of a polar disc does not resolve the riddle of the MW DoS because in addition to the polar material, the whole population of dwarf galaxies sitting within dark matter subhalos would still be present. As \citet{Kroupa2010} note, for satellites having independent infall and accretion histories, \citet{Libeskind2009} have shown that ''the DM hypothesis is much less likely than 0.4 per cent to be able to account for the MW satellite system in MW-type DM halos''. A result that has been obtained by only taking into account the phase-space distribution of the 11 classical satellite galaxies. The probability is therefore further diminished by the finding of additional systems aligned with the DoS. \textit{Individually} infalling dark matter subhalos do not align in a thin disc. The satellite galaxy system can therefore not be explained with any \textit{significant} confidence as being dark matter subhalos with \textit{independent} fall-in histories.

To be nevertheless explained as dark-matter objects, the common phase-space distribution of the MW satellites is often understood as evidence of their common origin. Along this line of reasoning, several attempts to explain the DoS were put forward within the CDM paradigm, most interpreting it as a cosmological accretion structure of DM dominated subhalos. The satellites might have formed in a common group of subhalos or have fallen in along the same filament.
These scenarios are not without problems, either. One suggestion is that the satellite galaxies fell into the MW halo as a group \citep{LiHelmi2008, DOnghiaLake2008}. However, the DoS is too thin to be compatible with the size of known associations of dwarf galaxies \citep{Metz2009b}. In addition, the scenario lacks in internal consistency: on the one hand, \citet{Deason2011} find that, to explain the coherency in velocity space, the group infall has to have happened recently. The only example of a group infall the authors show in their models happened at a redshift of 0.6. On the other hand, \citet{Nichols2011}, ignoring the spatial anisotropy, find that to explain the radial distribution of gas-deficient and gas-rich dwarfs around the MW, the galaxies had to have fallen in early (at a redshift of 3-10).

A second attempt \citep{Lovell2011} suggests that filamentary accretion of the satellite galaxies results in a quasi-planar distribution of coherently rotating satellites. However, according to fig. 1 in \citet{Lovell2011} only a maximum of about 22 (35) per cent of the subhalos\footnote{In case of isotropy this would be 10 (20) per cent.} would have an orbital pole within $26^\circ$\ ($37^\circ$) of the main halo spin axis, which at best only barely qualifies as a coherently rotating satellite system. More importantly, the authors themselves explain that the satellites would preferentially orbit within the plane of the MW galaxy. In high-resolution simulations of galaxies forming in dark matter halos, \citet{Bett2010} find that it is most likely for a galaxy to have its spin vector aligned with the direction of the angular momentum of the parent halo. The median misalignment between the two vectors is only $\approx 30^\circ$. \citet{Hahn2010} also find this orientation to be most likely in a simulation of galaxies forming in a cosmic filament, but report a larger median misalignment angle of $\approx 50^\circ$. They speculate that this discrepancy compared to the result of \citet{Bett2010} might originate from resolution issues or the particular feedback recipes employed. Another possible origin are the different large-scale environments of the simulations, the galaxies of \citet{Bett2010} might reside in a lower density environment than the filament of \citet{Hahn2010} and have had a more quiescent formation history, which might better resemble the quiescent merger history of the MW.

The DoS is oriented polar to the MW disc, so almost perpendicular to the most probable orientation. This whole reasoning, however, is rendered obsolete by recalling that in addition to the at most 22 (35) per cent alleged coherently orbiting, quasi-planar distributed DM halos in \citet{Lovell2011}, another 78 (65) per cent subhalos are present in the main halo outside of this distribution, which should lead to about four (two) times as many satellite galaxies outside the quasi-planar distribution of the DoS than inside. These are not observed around the MW. Furthermore, \citet*{Angus2011}, using the proper motions of the MW satellites, show that infall is not viable because the orbital angular momentum can only be dissipated through dynamical friction on the MW dark matter halo, implying the satellite galaxies to be hosted by unphysically massive dark matter halos. Thus, however one turns the coin, there is no consistent solution to the DoS problem in contemporary popular cosmology.

In opposition to the dark-matter-motivated accretion scenario, the DoS galaxies might be of tidal origin, having been formed in a galaxy collision \citep{Kroupa2010}. This scenario can be considered to be much more natural. The existence of the bulge of the MW and its formation on a very short timescale \citep{Ballero2007} already independently indicates that the young MW experienced an interaction, and at least partial merger, with another galaxy.
The tidal scenario would make the MW satellite galaxies mostly dark-matter free ancient Tidal Dwarf Galaxies (TDGs) formed in an interaction of the MW progenitor and another galaxy. Their positions and orbits would then consistently lie in the plane of the interaction and counter-orbiting satellites emerge naturally \citep*{Pawlowski2011}. Furthermore, in tidal interactions not only TDGs are produced, but also large numbers of massive star clusters with masses of $10^5$\ to $10^7 M_{\sun}$ \citep*{Bournaud2008}. These would have to be distributed around the MW in a similar manner.

A purely spatial alignment of MW satellites might be a (however unlikely) chance result of the time of observation. Finding a similar disc-like distribution in another set of supposedly accreted objects, globular clusters classified as young halo (YH) objects, would decrease the probability of the DoS being a non-genuine structure even more. In the tidal scenario, YH objects would have been formed together with, and thus would be similarly distributed as, the satellite galaxies in the DoS.

An alignment of the orbital planes of the non-satellite-galaxy-objects with the DoS would further support its significance, while a contrary result would cast doubt on the scenario of a pure tidal origin of the DoS. Therefore, more information on the orbits of MW satellites is needed. One way to add more orbit-information is to look at the disrupted remnants of clusters and dwarf galaxies around the MW: stellar and gaseous streams. Their benefit is that the spatial extend of each of these structures already defines an orbital plane for the respective stream. Thus, no additional, usually rather uncertain proper motion data are needed. Until further and more precise proper-motions for MW satellite galaxies are available, this method thus is a promising alternative in constraining not only the spatial, but also the kinematic anisotropy of the MW-halo substructure.

The distribution of globular clusters is analysed in Section \ref{sect:clusters}. In Section \ref{sect:streams}, at first the method for finding the stream normal vectors is presented (Section \ref{sect:method}) before individual streams are considered (Section \ref{sect:streamdata}). This is followed by a discussion of the preferential direction found (Section \ref{sect:results}) and an estimation of the probability of such a distribution of stream normals in case of an underlying isotropic distribution (Section \ref{sect:probability}), ending in a discussion of the effect of the stream-distance on the analysis (Section \ref{sect:anchorpointdistanceeffects}). Section \ref{sect:discussion} consists of our discussion of the discovered vast polar structure (VPOS) and the proposed scenario of the MW satellites being tidal dwarf galaxies. Finally, our conclusions are given in Section \ref{sect:conclusions}.

\section{Globular Cluster Distributions}
\label{sect:clusters}

In the following, the globular cluster (GC) system of the MW is analysed for a preferred spatial alignment with the DoS. For this, the GCs are split up into three groups. The old halo (OH) and bulge/disc (BD) objects are thought to have formed with the early MW, the latter being confined to the MW bulge and disc. A significant fraction of the young halo (YH) objects is found to be younger than the OH clusters.
They are thought to be of accretion origin \citep{Zinn1993, Parmentier2000}. This makes them interesting in the context of the DoS as, when they are of similar origin non-native to the MW, they can be expected to show an imprint of this origin in their spatial distribution like the MW satellite galaxies.

The GCs are divided into these three subsystems according to their metallicity and horizontal-branch morphology index. For the present analysis, the classification into horizontal-branch classes by \citet{Mackey05} is used. This results in sample sizes of 30 YH, 70 OH and 37 BD globular clusters. The cluster positions are taken from the catalog by \citet{Harris1996} (2003 update). 

To each sample of GCs a best-fitting plane is determined and compared to the DoS. The same method as described in section 5 of \citet{Kroupa2010} to determine the normal vector to the best-fitting disc of the satellite galaxies is now applied to the GCs of the MW. In short, the GC positions are expressed in Galactic, Cartesian coordinates. Starting with a projection onto the x-z plane and rotating it around the z-axis (pointing to the MW north pole), for each projection a best-fitting line is determined, the line being a plane seen edge-on in the current projection. The fit parameters of the projection with the lowest RMS height of GCs from the line-fit are then adopted, resulting in the direction of the normal to the best-fitting GC disc.

The YH clusters individually exhibit similarities to the star clusters of satellite galaxies. If the YH objects are of a similar origin as the MW satellite galaxies, then their best-fitting disc should consequently show similarities to the DoS, while these should be absent in the OH and BD populations. The BD clusters however should correlate with the plane of the MW. The inclusion of the BD clusters thus acts as a consistency check, their normal has to be  close to the MW pole. The analysis of the OH cluster sample can act as a test whether the found substructure is real. If observational biases would play a role, the OH clusters should be similarly affected and thus the analysis should find a similar normal vector for them.

\begin{figure*}
 \centering
 \includegraphics[width=180mm]{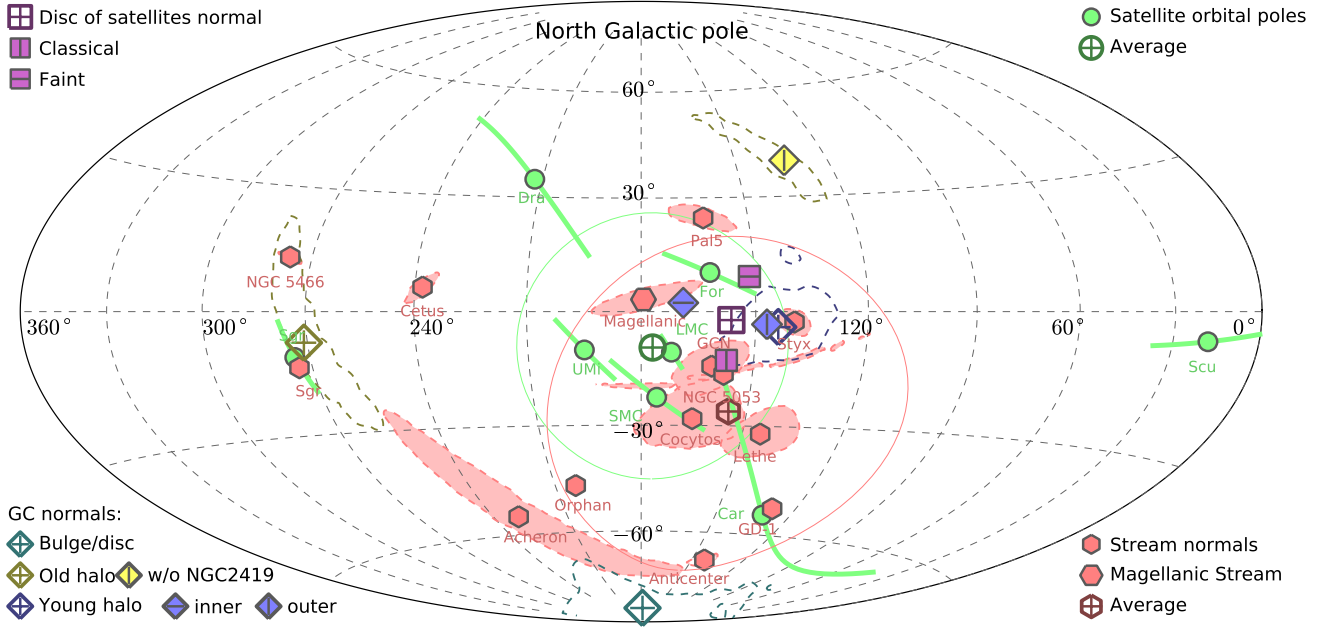}
 \caption{ Directions of normal vectors and orbital poles in Galactocentric coordinates; $l_{\rm{MW}}$\ is the Galactic longitude, $b_{\rm{MW}}$\ the Galactic latitude. Viewed from the Galactic centre, the Sun is at $l_{\rm{MW}} = 180^\circ$, $b_{\rm{MW}} = 0^\circ$.
  For a better readability only one of the two normals for each object is shown (those in the range $120^\circ < l_{\rm{MW}} < 300^\circ$).
 \textbf{Disc of Satellites:} The filled, light magenta squares show the normal vectors for the DoS fitted to the 11 classical (vertical line) and 13 faint (horizontal line) satellite galaxies from \citet{Kroupa2010}. The fit to all 24 satellite galaxies together gives the normal vector illustrated by the open, dark magenta square. 
 \textbf{Satellite orbital poles:} The orbital poles of the eight MW satellites analysed by \citet{Metz2008} are included as light-green dots. The solid light-green lines show the uncertainties in the orbital pole positions. The larger, open, green circle marks the average orbital pole of the six satellite galaxies that lie close to the DoS, while the thin green circle gives their spherical standard deviation of $35.4^\circ$.
 \textbf{Globular clusters:} The normal vector of the disc fitted to the globular clusters (see Sect. \ref{sect:clusters}) classified as young halo objects, DoGC$_{\rm{YH}}$, is included as the open blue diamond, that for the old halo clusters, DoGC$_{\rm{OH}}$, as an open yellow diamond and that of the bulge/disc clusters, DoGC$_{\rm{BD}}$, as an open cyan diamond. The dashed contours in the respective colors represent the 1-$\sigma$\ regions for the normal vectors derived from bootstrapping. They, too, are shown in only one direction, within $120^\circ < l_{\rm{MW}} < 300^\circ$\ for the YH and OH clusters and only at the southern MW pole for the BD clusters. The yellow filled diamond denotes the normal vector to the OH objects when excluding the outlying cluster NGC 2419. The filled blue diamond with a horizontal line marks the direction of the normal vector fitted to the 20 inner YH clusters (Galactocentric distances less than 20 kpc), while the one with a vertical line shows the normal direction fitted to the outer YH clusters only (Galactocentric distances of more than 25 kpc).
 \textbf{Streams:} The filled, light-red hexagons denote the directions of the stream-normal vectors (see Sect. \ref{sect:streams}) derived from the stream anchor points compiled in Table \ref{tab:streamdata}. The Magellanic Stream-normal is marked by slightly bigger, rotated hexagon. The uncertainties derived from Monte-Carlo experiments are illustrated as the light-red areas surrounded by dashed 1-$\sigma$ contours. The open, dark-red hexagon represents the direction of the average stream-normal, its spherical standard distance is illustrated with the thin red circle.
 There is a strong clustering close to the DoS normal, indicative of the vast polar structure (VPOS) around the MW. The majority of satellite galaxy orbital poles, half the stream-normal vectors and the DoGC$_{\rm{YH}}$\ normal vector all lie in the vicinity of the DoS normal. There is very good agreement of the DoS normal and the average stream-normal directions and they both are close to the mean orbital pole. The DoS normal lies within the standard deviations of the streams, orbital poles and the DoGC$_{\rm{YH}}$\ normal vector.
 The DoGC$_{\rm{OH}}$\ and DoGC$_{\rm{BD}}$\ normals are obviously detached, with the latter aligning with the MW pole as expected.\label{fig:normalvectorsbig}}
\end{figure*}

This analysis does not quantify the significance of the prolate-ness of the GC distributions, but merely determines the distributions' minor axes at each projection. To test the robustness of the determined normal vectors, a bootstrapping analysis is performed. For each of the three globular cluster subsystems, the analysis is repeated 10000 times on a randomly re-sampled cluster population of the same sample size, allowing for multiple occurrences of the same clusters. The distribution of the normal vectors produced by this method is illustrated by the $1\sigma$\ contours in Fig. \ref{fig:normalvectorsbig}. Furthermore, the mean directions and spherical standard deviations from these have been determined. Finally, the shape-parameter $\gamma$\ and strength-parameter $\zeta$\ \citep{Fisher1987, Metz2007} allow to asses whether the distribution of bootstrapped normal vectors is clustered or girdled ($\gamma > 1$ and $\gamma < 1$, respectively), and how strongly so (the larger $\zeta$, the more concentrated the distribution is, with an isotropic distribution having $\zeta = 0$).
The resulting normal vectors are compiled in Table \ref{tab:normalpositions}\footnote{
Note that, independently and in parallel, \citet*{Keller2011} have also analysed the globular cluster system of the MW and report similar results, even though their GC categorization leads to slightly different samples. Their full YH clusters result in a normal vector of the best-fitting plane pointing to $(l, b) = (144^{\circ} \pm 6^{\circ}, -13^{\circ} \pm 5^{\circ})$, while fitting the outer 7 YH clusters only leads to $(l, b) = (162^{\circ} \pm 5^{\circ}, -8^{\circ} \pm 7^{\circ})$.}.

\begin{table*}
\begin{minipage}{180mm}
 \caption{Directions of normal-vectors}
 \label{tab:normalpositions}
 \begin{center}
 \begin{tabular}{@{}lccccll}
  \hline
    Name & Symbol in Fig. \ref{fig:normalvectorsbig} & Method & $l [^\circ]$ & 
  $b [^\circ]$ & Objects & Reference \\
  \hline
  DoS 2007 & square (vertical) & positions & 157.3 & -12.7 & 11 classical satellites & \cite{Metz2007} \\
  DoS 2009 & not plotted & positions & 149.6 & -5.3 & 22 satellites & \cite{Metz2009} \\
  DoS 2009 (w/o Her) & not plotted & positions & 159.7 & -6.8 & 21 satellites & \cite{Metz2009} \\
  DoS & open square & positions & 156.4 & -2.2 & 24 satellites & \cite{Kroupa2010} \\
  Faint DoS & square (horizontal) & positions & 151.4 & 9.1 & 13 non-classical satellites& \cite{Kroupa2010} \\
  Mean orbital pole & open dot & proper motions & 177.0 & -9.4 & 6 satellites& \cite{Metz2008} \\
  Average stream-normal & open hexagon & anchor points & 155 & -26 & 14 streams & this work \\
  DoGC$_{\rm{YH}}$ & blue diamond & positions & 144 & -4.3 & 30 GC classified as YH & this work \\    
  DoGC$_{\rm{YH}}$ ($< 20~\rm{kpc}$) & blue diamond (horizontal) & positions & 169 & 2.3 & 20 inner YH GCs & this work \\    
  DoGC$_{\rm{YH}}$ ($> 25~\rm{kpc}$) & blue diamond (vertical) & positions & 147 & -3.3 & 10 outer YH GCs & this work \\    
  DoGC$_{\rm{OH}}$ & yellow diamond & positions & 271 & -7.4 & 70 GC classified as OH & this work \\
  DoGC$_{\rm{OH}}$ (w/o NGC 2419) & yellow diamond (vertical) & positions & 134 & 39.6 & 69 GC classified as OH & this work \\
  DoGC$_{\rm{BD}}$ & cyan diamond & positions & 175 & -85.7 & 37 GC classified as BD & this work \\
  \hline
 \end{tabular}
 \end{center}
 \small \medskip
Collection of the directions of the normal vectors of planar structures around the MW, given in Galactocentric, Galactic longitude $l$\ and latitude $b$. Listed are DoS- and DoGC-fit-normals, the average satellite orbital angular momentum and the average stream-normal position. Directions of individual stream-normals can be found in Table \ref{tab:streamdata}. Note the close proximity of results obtained with independent methods and objects: satellite galaxy positions and proper motions, stream-normals and YH globular cluster positions follow the same VPOS.
\end{minipage}
\end{table*}

\subsection{Young Halo GCs}
For the YH clusters (blue open diamond in Fig. \ref{fig:normalvectorsbig}), a striking agreement with the DoS of the MW satellite-galaxies is found. The 'disc of young halo globular clusters' (DoGC$_{\rm{YH}}$) normal points towards $(l, b) = (144^\circ, -4.3^\circ)$, less than $13^\circ$\ away from the DoS normal. The chance that a randomly oriented vector is this close to the DoS normal is only about 2.5 per cent. The root-mean-square (RMS) scale height of GCs from the best-fitting plane is 11.8 kpc, its offset from the Galactic centre is 2.6 kpc. 
The bootstrapping analysis finds a mean direction of $(l, b) = (145.2^\circ, -1.3^\circ)$\ with standard deviation of $21.5^\circ$\ which is consistent with the DoGC$_{\rm{YH}}$-direction. The shape- and strength parameters are $\gamma = 3.12$\ and $\zeta = 3.36$, i.e. a significant clustered distribution of bootstrapped normal vectors. Taken together it can be stated that the YH globular cluster distribution is indistinguishable from that of the MW satellite galaxies. Both show a preference to lie in the same plane highly inclined to the MW disc. 

The fitting method is dominated by the outermost GCs. There is a natural radial separation point in the sample, as the 10 outermost YH clusters have Galactocentric distances of more than 25 kpc, while the inner 20 are closer than 20 kpc. To test the domination of the 10 outermost YH clusters, those were fitted individually. The fit gives essentially the same normal vector as the full sample: $(l, b) = (147^\circ, -3.3^\circ)$, with an angular distance from the DoS normal of now only $9.5^\circ$\ and an RMS scale height of 18.3 kpc. 
The offset of the best-fitting plane for this sample from the Galactic centre is 8.0 kpc, almost the same as the offset of 8.2 kpc found by \citet{Kroupa2010} for the 24 MW satellites. Both offsets point into the same direction (towards the reported normal vectors, i.e. away from the Galactic centre, approximately in the anti-centre direction), emphasizing the common distribution of YH GCs and MW satellite galaxies. Interestingly, the planes are shifted into the approximate direction of the Andromeda Galaxy, which lies at galactic coordinates $(l,b) = (121.2^\circ, -21.6^\circ)$. This is $39^\circ$\ away from the DoS normal and $29^\circ$\ away from the normal defining the best-fitting plane to the outer 10 YH GCs.

A fit to the remaining inner 20 YH GCs gives a normal vector of $(l, b) = (169^\circ, 2.3^\circ)$, $13.4^\circ$\ away from the DoS normal, with a RMS scale height of 7.3 kpc. While the two normals to subsets of the YH objects are $22.7^\circ$\ apart from each other, they lie on opposing sides of the DoS normal direction and thus are both aligned very close to the DoS.

This suggests that the DoS is a structure that has left an imprint at least as close as 20 kpc from the MW centre. The probability to find two independent normal vectors drawn from a uniform distribution within $13.4^\circ$\ of the DoS is less than 0.1 per cent, a strong hint that the two YH GC samples share a common information in their distribution.

\subsection{Old Halo GCs}
The best fitting normal vector found for the OH clusters DoGC$_{\rm{OH}}$\ (yellow open diamond in Fig. \ref{fig:normalvectorsbig}) points to $(l, b) = (271^\circ, -7.4^\circ)$\ with a RMS scale height of 4.8 kpc and an offset from the Galactic centre of 1.8 kpc. This, while also highly inclined to the MW disc, is $66^\circ$\ away from the DoS normal. However, the DoGC$_{\rm{OH}}$\ is strongly dominated by the outermost cluster, NGC 2419, which has a Galactocentric distance of about 90 kpc. The second-farthest cluster, Palomar 15, is at less than 40 kpc distance. Excluding NGC 2419, the resulting normal vector points to $(l, b) = (134^\circ, 39.6^\circ)$, still $47^\circ$\  away from the DoS, and results in a RMS scale height of 3.9 kpc. The OH globular cluster distribution is therefore not aligned with the DoS. The bootstrapping analysis further shows that the evidence for a best-fitting disc for the OH GCs is less convincing. The bootstrapped normal vectors of the full sample show a larger scatter with a standard deviation of $30.7^\circ$\ ($23.2^\circ$\ excluding NGC 2419). Their mean normal vector points to $(l, b) = (279.3^\circ, -15.8^\circ)$\ ($(133.1^\circ, 40.2^\circ)$\ excluding NGC 2419). The angular distance to the normal vector of the non-bootstrapped sample of all OH GCs is $12^\circ$. The shape- and strength-parameters $\gamma = 0.96$\ and $\zeta = 2.93$\ ($\gamma = 1.32$\ and $\zeta = 3.62$) are not conclusive either, being on the edge between a clustered $(\gamma > 1$) and a girdled ($\gamma < 1$) distribution. Interestingly, the resulting normal vector of the sample including all OH GCs points close to the orbital pole of the Sagittarius dwarf galaxy at $(l, b) = (273^\circ, -13.5^\circ)$.

\subsection{Bulge/Disc GCs}
The BD clusters behave as expected. The analysis finds a best-fitting DoGC$_{\rm{BD}}$\ normal vector (cyan open diamond in Fig. \ref{fig:normalvectorsbig}) pointing towards $(l, b) = (175^\circ, -85.7^\circ)$, very close to the MW pole. The RMS scale height is 1.3 kpc, the offset from the Galactic centre is 0.2 kpc. The bootstrapping confirms this direction, with a mean bootstrapped normal vector at $(l, b) = (166.3^\circ, -87.5^\circ)$\ and a standard deviation of only $12.7^\circ$. The shape- and strength parameters of $\gamma = 2.95$\ and $\zeta = 4.40$ indicate further that the bootstrapped normal vectors are strongly clustered. The BD clusters agree well with forming a plane co-planar to the MW disc, a result which gives confidence in the informative value of the analysis.

These findings thus support the validity of the DoS as a major structural component of the MW. The vastly differing plane normal directions for the OH and the YH clusters and the less well-defined properties of the former confidently demonstrate that the agreement in normal directions of the DoS and the DoGC$_{\rm{YH}}$\ are not merely based on some unknown bias in the data.

\section{Streams}
\label{sect:streams}

\subsection{Method}
\label{sect:method}

The assumption that stars in tidal streams approximately follow the orbit of their parent object \citep{Odenkirchen2003} is supported by numerical simulations for both globular clusters \citep*{Combes1999} and dSph satellite galaxies \citep*{Johnston1996, Zhao1999}. \citet{Montuori2007} found that, close to a cluster (within 7-8 tidal radii), the tidal tail mainly point towards the Galactic centre and thus is a poor tracer of the cluster path. Longer tails, however, are good tracers of the cluster path, which is the case for the extended streams analysed in this work. For very massive satellite galaxies, it has been shown \citep*{Choi2007} that the tidal tails can deviate strongly from the path of the satellite. This deviation, however, is in the radial direction as seen from the Galactic centre, which does not affect our analysis of the stream-normal vectors because the tail remains confined in the orbital plane of the satellite galaxy (see their fig. 7).

To find the stream-normal vectors, two points along the stream are chosen. These are, depending on the available data in the literature, the reported start- and end-points of streams or pronounced over-densities along the streams. The stream-normal vectors are then determined to be the normal vectors of a plane defined by these two anchor points and the centre of the MW, thus assuming the streams' centre of orbit is the MW centre. This assumption is warranted as the MW centre should be the centre of mass of the Galaxy.

\begin{figure}
 \centering
 \includegraphics[width=80mm]{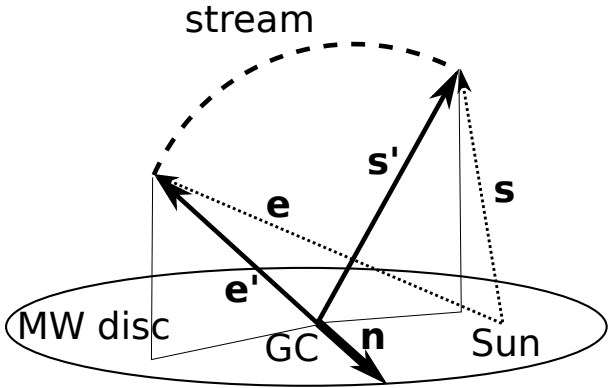}
 \caption{Sketch illustrating the determination of the stream-normal vectors. The stream is illustrated by the curved, dashed line. The vectors $\mathbf{s}$\ and $\mathbf{e}$\ are pointing from the position of the Sun towards the start- and the end-anchor-point of the stream, respectively. They are transformed into the vectors $\mathbf{s'}$\ and $\mathbf{e'}$\ which point from the Galactic centre (GC) torwards the stream's start- and end-point, respectively. Their cross product results in the stream-normal vector $\mathbf{n}$.}
 \label{fig:vectorsketch}
\end{figure}

With these assumptions, the position of the normal to the plane of the stream is estimated. The two anchor points on a stream are $\mathbf{s}$\ and $\mathbf{e}$. These are two vectors pointing from the Sun to two parts of the stream (see Fig. \ref{fig:vectorsketch}), given in Galactic, Heliocentric coordinates in Table \ref{tab:streamdata}. In the first step, the vectors $\mathbf{s}$\ and $\mathbf{e}$\ are transformed to Galactic, Cartesian coordinates. They are then shifted by $d_{\rm{GC}}$, the distance of the Sun from the Galactic centre, along the x-axis. For consistency with \citet{Metz2007} $d_{\rm{GC}} = 8.5~\rm{kpc}$\ is chosen. This places the origin of the Cartesian coordinate system for the transformed vectors $\mathbf{s}'$\ and $\mathbf{e}'$\ on the Galactic centre. The \textit{stream-normal vector} $\mathbf{n}$\ to the plane defined by the two points on the stream and the Galactic centre becomes
\begin{displaymath}
\mathbf{n} = \frac{\mathbf{s}' \times \mathbf{e}'}{|\mathbf{s}' \times \mathbf{e}'|}.
\end{displaymath}

The stream-normal vector will then be expressed in Galactocentric coordinates of longitude $l$\ and latitude $b$, defining its direction.

This method is easy to implement and can be used with available data for streams from the literature. The major sources of error are the uncertain distances to the two anchor points and the distance to the centre of the MW. Furthermore, in cases of extremely diffuse streams spreading over a large angular scale, the angular position of the anchor points can be badly defined. To estimate these uncertainties, we adopt a Monte Carlo method. We vary the anchor-point angular positions and radial distances according to a Gaussian distribution with FWHM of the uncertainties as compiled in Table \ref{tab:streamdata}. The resulting uncertainty in the position of the stream-normal vector for 10000 Monte-Carlo realisations is illustrated by 1-$\sigma$\ contours in Fig. \ref{fig:normalvectorsbig}. 

For each stream-normal vector, the angular distance $\Theta$\ from the DoS normal vector pointing to $l = 156.4^\circ$\ and $b = -2.2^\circ$\ \citep{Kroupa2010} is calculated. The uncertainties in this angular distance are not necessarily symmetric. An extreme example is the stream of NGC 5053, which has uncertainty-contours that are almost tangential to the DoS normal and thus more extended to higher angular distances from the DoS normal than to lower. Therefore, the uncertainties are expressed as follows: the lower (upper) uncertainty interval is that interval within which fall $1 \sigma$\ of the Monte-Carlo realisations that lead to smaller (larger) distances from the DoS normal than the non-varied case. 
The results are listed in Table \ref{tab:streamdata}.

\begin{table*}
\begin{minipage}{180mm}
 \caption{Streams of the Milky Way}
 \label{tab:streamdata}
 \begin{center}
 \begin{tabular}{lcccccccccl}
  \hline
    Name & Type & $l^{\rm{HC}} [^\circ]$ & $b^{\rm{HC}} [^\circ]$ &  $\Delta lb [^\circ]$ & $r^{\rm{HC}} [\rm{kpc}]$ & $\delta [^\circ]$ & $l_{\rm{N}}^{\rm{GC}} [^\circ]$ & $b_{\rm{N}}^{\rm{GC}} [^\circ]$ & $\Theta [^\circ]$ & ref \\
  \hline
  GCN stream & GS & 39.3 & -31.0 & $\pm 1$ & $25 \pm 5$ & 80 & 161 & -14 & $13 ^{+7}_{-4}$ & (7) \\
 & & 18 & 47 & $\pm 3$ & $11 \pm 1$ &  &  &  &  &  \\
NGC 5053 stream & GC & 312.7 & 77.7 & $\pm 0$ & $16.4$ & 5 & 158 & -16 & $14 ^{+21}_{-1}$ & (8) \\
 & & 335.6 & 79.0 & $\pm 1$ & $16.4 \pm 1.64$ &  &  &  &  &  \\
Styx & DG & 311.5 & 82.8 & $\pm 1.7$ & $38.0 \pm 7.6$ & 60 & 140 & -3 & $17 \pm 3$ & (3) \\
 & & 42.7 & 30.0 & $\pm 1.7$ & $50 \pm 10$ &  &  &  &  &  \\
Magellanic Stream & DG/GS & 73.4 & -68.0 & 2.0 & $55.0 \pm 11.0$ & 18 & 179 & 3 & $24 ^{+10}_{-8}$ & (1) \\
 & & 29.4 & -83.4 & $\pm 2.0$ & $55.0 \pm 11.0$ &  &  &  &  &  \\
Cocytos & GC & 289.5 & 59.2 & $\pm 1$ & $11 \pm 2$& 75 & 165 & -28 & $27 ^{+9}_{-6}$ & (3) \\
 & & 41.6 & 29.6 & $\pm 1$ & $11 \pm 2$ &  &  &  &  &  \\
Pal 5 stream & GC & 355.3 & 45.9 & $\pm 0.5$ & $23.2 \pm 2.0$ & 22 & 163 & 25 & $27 ^{+4}_{-3}$ & (4) \\
 & & 22.3 & 36.3 & $\pm 0.5$ & $23.2 \pm 2.0$ &  &  &  &  &  \\
Lethe & GC & 232.7 & 67.9 & $\pm 0.5$ & $12.2 \pm 2.4$ & 81 & 145 & -32 & $32 ^{+5}_{-3}$ & (3) \\
 & & 41.2 & 30.5 & $\pm 0.5$ & $13.4 \pm 2.6$ &  &  &  &  &  \\
GD-1 & GC & 196.7 & 47.4 & $\pm 0.25$ & $7.7 \pm 0.5$ & 48 & 129 & -52.3 & $55 \pm 1$ & (5) \\
 & & 113.9 & 58.1 & $\pm 0.25$ & $9.1 \pm 0.5$ &  &  &  &  &  \\
Orphan & DG & 249.5 & 50.0 & $\pm 0.7$ & $21.4 \pm 1.0$ & 49 & 203 & -47 & $60 ^{+2}_{-1}$ & (11) \\
 & & 173 & 46.5 & $\pm 0.7$ & $46.8 \pm 4.5$ &  &  &  &  &  \\
NGC 5466 stream& GC & 42.2 & 73.6 & $\pm 0.0$ & $16.3 \pm 0.0$ & 29 & 276 & 13 & $60 \pm 4$ & (6) \\
 & & 154.9 & 71.8 & $\pm 2$  & $16.3 \pm 3.3$ &  &  &  &  &  \\
Sagittarius & DG & ... & ... & ...  & ... & ... & 273.8 & -13.5 & 64 & (9) \\
Anticentre & DG & 224.8 & 20.5 & $\pm 2.5$ & $8.9 \pm 0.2$ & 65 & 143 & -68 & $67 \pm 1$ & (2) \\
 & & 151.0 & 37.6 & $\pm 2.5$ & $8.9 \pm 0.2$ &  &  &  &  &  \\
Acheron & GC & 359.8 & 43.9 & $\pm 0.5$ & $3.8 \pm 0.76$ & 37 & 230 & -55 & $79 \pm 8$ & (3) \\
 & & 42.7 & 30.0 & $\pm 0.5$ & $3.5 \pm 0.70$ &  &  &  &  &  \\
Cetus Polar Stream & GC/DG & 144 & -71 & $\pm 2$ & $36.1 \pm 1.9$ & 25 & 238 & 6 & $82 \pm 3$ & (10) \\
 & & 142 & -46 & $\pm 4$ & $30.1 \pm 1.6$ &  &  &  &  &  \\
  \hline
 \end{tabular}
 \end{center}
 \small \medskip
List of MW streams analysed in this work. Type lists the type or suspected origin of the stream, being either a gaseous stream (GS), a stellar stream of globular cluster (GC) or of dwarf galaxy (DG) origin. For each stream, the Heliocentric coordinates of two anchor points are listed, with Galactic longitude $l^{\rm{HC}}$, Galactic latitude $b^{\rm{HC}}$, angular uncertainty $\Delta lb [^\circ]$\ and Heliocentric distance $r^{\rm{HC}}$. These have been extracted from the papers listed under ref, as described in Section \ref{sect:streamdata}. The respective uncertainties are adopted from these sources or have been estimated. The Sagitarrius stream is an exception to this, its stream-normal is directly taken from \citet{Majewski2003}. The angle $\delta$\ is the angular distance between the two anchor points. The resulting stream-normal vectors point to Galactic longitude $l_{\rm{N}}^{\rm{GC}}$\ and latitude $b_{\rm{N}}^{\rm{GC}}$, this time in Galactocentric coordinates. The angular distance from the DoS normal vector is $\Theta$. \\
References: (1) \citet{Bruens2005}; (2) \citet{Grillmair2006b}; (3) \citet{Grillmair2009}; (4) \citet{GrillmairDionatos2006a}; (5) \citet{GrillmairDionatos2006b}; (6) \citet{GrillmairJohnson2006}; (7) \citet{Jin2010}; (8) \citet{Lauchner2006}; (9) \citet{Majewski2003}; (10) \citet{Newberg2009}; (11) \citet{Newberg2010}.
\end{minipage}
\end{table*}

\subsection{Individual stream data}
\label{sect:streamdata}

The method described is applied to 14 streams (stellar and gaseous) in the MW halo for which anchor points with angular distances of more than $5^\circ$\ could be extracted from the literature. No shorter streams or mere tidal deformations were taken into account. The individual streams, the choice of anchor points and possible problems or difficulties are discussed below. The stream data used are compiled in Table \ref{tab:streamdata}, together with the respective references.
In total, all known\footnote{To the authors.} (and suspected) stream-structures close to the MW are included in the analysis. Reasons for leaving out the Virgo Stellar Stream (VSS) from the present analysis are given at the end of this section.

The streams can be divided into two groups, one having stream-normal vectors pointing close to the DoS normal (less than $35^\circ$\ away) and those further away (more than $50^\circ$). In the following, they are sorted by their stream-normal vector's angular distance from the DoS normal, starting with the closest stream.

\textbf{GCN}: The GCN high velocity cloud complex shows a string of HI clouds which \citet{Jin2010} show to probably be a gaseous stream. The author fits an orbit to a number of GCN clouds. Jin's initial-condition point for the orbit estimation is taken as one anchor point for the stream. Its radial distance results from the orbit calculation. The second anchor point is the position of a compact high velocity cloud which Jin finds to be close to the orbit estimate in position and radial velocity. The radial distance to this second anchor point is estimated from the radial distance to the orbit fitted by \citet{Jin2010} (from her figs. 3 and 4), which does not vary much in this part of the suspected stream. 

The GCN high velocity cloud complex is the structure which has a stream-normal closest to the DoS normal of all streams analysed in this work, with an angular distance of only $13^{\circ}$. But the object is also the most speculative one. Comparing locations and velocities of MW satellites with the GCN, no possible progenitor for the stream has been found. It may be an ancient remnant of the gaseous tidal tail which formed the satellites and YH GCs.

\textbf{NGC 5053}: Analysing SDSS data, \citet*{Lauchner2006} report the finding of a stellar tidal debris stream of approximately 1.7 kpc in projected total length, extending in one direction from the cluster NGC 5053. This is 26 times the cluster's tidal radius \citep{Mackey05}, therefore it can be expected that the tidal tail traces the orbit of the cluster. However, no continuous tail is detected, but a string of overdensities. This might indicate that the underlying stream is too faint to be detected such that only the overdensities of the tail are seen. Those might arise as epicyclic overdensities in the tail \citep{Kuepper2010}.
The indications of a tidal tail close to the cluster was confirmed by \citet{Jordi2010}, but they did not reproduce the larger structure.
One anchor point is chosen to be the cluster itself, with position and distance data adopted from the catalogue by \citet{Harris1996}. The second anchor point is the reported end of the suspected tidal tail \citep{Lauchner2006} assuming its Heliocentric distance to be 16.4 kpc, the same as the cluster. Uncertainties are only adopted for this second anchor point, $\pm 1^\circ$\ in position and 10 per cent in distance. 
The latter results in a radial uncertainty similar to the projected length, the stream's orientation is thus varied by about $\pm 45^\circ$. As the radial velocity of NGC 5053 in the Galactic standard of rest is only 34 km/s \citep{Law2010}, it is reasonable to assume that the orbit is not aligned closely with the line of sight and that the tail therefore is seen more from its side than along its extend.

The stream-normal for the NGC 5053 stream points close to the DoS normal, with an angular distance of $14^{\circ}$. Thus, if this stream indeed follows the orbit, the cluster can not be associated with the Sagittarius dwarf galaxy, as stated by \citet{Law2010}, because the orbits of Sagittarius and NGC 5053 would be almost perpendicular to each other.

\textbf{Styx}: The origin of this stellar stream is most likely a disrupted dwarf galaxy. The reported start and end positions \citep{Grillmair2009} are adopted as anchor points, including the approximate distances. The width of the stream was used to estimate the angular uncertainty for the anchor points, the distance errors are estimated to be 20 per cent individually. A possible origin is the Bo\"otes III stellar over-density, also discovered by \citet{Grillmair2009}. Bo\"otes III was later analysed in detail \citep{Carlin2009}, leading to the conclusion that it is likely a currently disrupted dwarf galaxy in a transitional state between being a bound object and an unbound tidal stream \citep{Kroupa1997}.

The Styx stellar stream has a stream-normal which, with $17^{\circ}$\ distance, is the third-closest stream to the DoS normal. Bo\"otes III's position is only $\approx 8.5$~kpc away from the DoS plane. The excellent agreement of the stream-normal (and thus the orbit of the satellite) with the normal of the DoS supports therefore the claims by \citet{Metz2008} that the DoS is rotationally supported.

\textbf{Magellanic Stream}: The first anchor-point of this gaseous stream is chosen to be the position at which the two filaments of the stream overlap, point B in fig. 12 of \citet{Bruens2005}. The other anchor point is point C in their fig. 13, marking an over-density in the Stream. The distance estimate to the Magellanic Stream of 55 kpc is adopted, deduced from the average of the distances of the Large and Small Magellanic Clouds (LMC \& SMC). The distance-uncertainty for the Monte Carlo experiment is chosen to be 20 per cent, for the position-uncertainties $2^\circ$\ are adopted.

The resulting stream-normal points close (within $15^\circ$) to the LMC orbital pole. As can be expected from the orbital poles of the LMC and SMC, the Magellanic Stream's stream-normal is close to the DoS, too. Its angular distance from the DoS normal is $24^{\circ}$. It is slightly offset from both the LMC and SMC orbital poles towards the north.

\textbf{Cocytos}: This stellar stream has most likely a globular cluster origin. A start and an end-position are given in the discovery paper \citep{Grillmair2009}, which are used as anchor points. The position uncertainties are estimated to be $0.5^\circ$, as the measured width of the stream is reported to be less than one degree and the positions are given to one degree accuracy. The distance-estimate for the stream is 11 kpc, with an uncertainty of 2 kpc which is adopted individually for both anchor points.

Cocytos has a stream-normal close to the DoS normal, at an angular distance of $27^{\circ}$. The major part of the offset is in Galactic latitude, while the stream-normal is very close to the DoS in Galactic longitude.

\textbf{Palomar 5}: this globular cluster shows rather extended stellar tidal tails. The anchor points are the start- and end-points of the stream as reported by \citet{GrillmairDionatos2006a}, as they cover the stream over a wider extent than \citet{Odenkirchen2001}. The anchor point distances are adopted from their best-fit orbits and are the same as the distance to the cluster Palomar 5. They are allowed to vary by 2 kpc, while the positions are varied by $0.5^\circ$.

Palomar 5 can be associated to the DoS as its stream-normal is only $27^{\circ}$\ away from the DoS normal. The cluster is at a distance of 14.8 kpc from the DoS plane and is classified as a YH object.

\textbf{Lethe}: This stellar stream has most likely a globular cluster origin. A start and an end-position with distance-estimates are given in the discovery paper \citep{Grillmair2009}, which are used as anchor points. 
The random distance uncertainties are estimated in that paper to be 10 per cent and the systematic one 11 per cent in case of the Acheron stream. As the same method is used for the Lethe distance determination, a relative line-of-sight uncertainty of 20 per cent is adopted here. Note that this over-estimates the stream-normal vector uncertainty. The distances to the two anchor points are varied individually in the Monte-Carlo analysis, while the systematic part of the uncertainty would affect the distance to both anchor points in the same way. The positional uncertainty is $0.5^\circ$, as the stream is much thinner but the positions are only reported to one degree accuracy.

Lethe, like Cocytos, has a stream-normal close to the DoS normal, at a distance of $32^{\circ}$. Again, the major part of the offset is in Galactic latitude, while the normal is very close to the DoS in Galactic longitude.

\textbf{GD-1}: As anchor points two positions on the stream are used for which \citet{GrillmairDionatos2006b} give distance estimates. These are the midpoint of the north-eastern segment of the stellar stream, which the authors used as the fiducial point for the orbit estimation, and the densest portion of the stream.  The distance uncertainties are 0.5 kpc, the same as used by \citet{GrillmairDionatos2006b}, and the position uncertainties for both anchor points are half the estimated FWHM of the stream.

The GD-1 stream-normal is $55^{\circ}$\ away from the DoS normal. Interestingly it is very close to the orbital plane of the Carina dwarf galaxy. It seems not to be associated with that galaxy, though, as the estimated orbital parameters \citep{GrillmairDionatos2006b} confine it to within $18 \pm 2$~kpc, while the Carina dSph is at a distance of $\approx 100~\rm{kpc}$ \citep{Mateo1998}. Furthermore, no known GC can currently be associated with the stream.

\textbf{Orphan}: For the present analysis, the recently published new investigation of the Orphan Stream by \citet{Newberg2010} is used, instead of the discovery papers \citep{Grillmair2006a, Belokurov2007}. Their table 2 compiles distances and positions for different parts of increasing Galactic longitude on the stellar stream. The chosen anchor points are the first and the second-last entry in that table. The last entry was derived using data of only three stars and has therefore not been considered, as even \citet{Newberg2010} excluded it from their analysis.

This stream-normal is $60^{\circ}$ away from the DoS normal, but is close to the circle describing the standard deviation from the mean orbital pole \citep{Metz2008}.

\textbf{NGC 5466}: There is a stellar tail associated with the globular cluster NGC 5466 \citep{GrillmairJohnson2006}. Unfortunately, no distance estimate to the stream is given. Because of this, the first anchor point is chosen to be the cluster itself, for which the distance is known to be 16.3 kpc \citep{Harris1996}, and its position is not varied in the Monte Carlo experiment. The second anchor point is the end of the stream as reported in the caption of fig. 1 in \citet{GrillmairJohnson2006}. For this second anchor point, the same distance as for the cluster is used, with a distance uncertainty of 20 per cent. 

While NGC 5466's distance from the DoS is only $\approx 2.9$~kpc, its stream-normal is far from the DoS normal, at an angular distance of $60^{\circ}$. The cluster is classified as a young halo cluster (see Section \ref{sect:clusters}) and as such has most probably been accreted by the MW or been scattered out of the DoS \citep[cf.][]{Zhao1998}.

\textbf{Sagittarius}: The orbital plane of the Sagittarius stellar stream was analysed in great detail \citep{Majewski2003, Fellhauer2006} and is in very good agreement with the orbital pole of the satellite galaxy. We thus simply adopt the \citet{Majewski2003} stream-normal of $l = 273.8^{\circ}$, $b = -13.5^{\circ}$.

It has already been pointed out that the Sagittarius satellite galaxy is not orbiting within the DoS \citep{Metz2008}, the stream-normal is at an angular distance of $64^\circ$\ from the DoS normal. However, it is close to the DoGC$_{\rm{OH}}$\ normal vector. It might have been scattered on its present orbit \citep{Zhao1998}, a possibility we discuss in Sect. \ref{sect:MWsatsasTDGs}.

\textbf{Anticentre}: The start and end positions of the stellar stream from \citet{Grillmair2006b} are used as anchor points in the analysis. The reported average Heliocentric distance and its uncertainty are adopted for both anchor points. The position uncertainties are estimated to be half the average stream width of $5^\circ$.
Due to the width of the stream, the most likely origin is a dwarf galaxy. There seems to be substructure in the stream, with at least two more narrow components accompanying the wider central stream. It is speculated that these sub-streams might be the remnants of a globular cluster population of the original dwarf galaxy.

The Anticentre stream, while its stream-normal's Galactic longitude is close to the DoS's one, is $67^{\circ}$\ away in Galactic latitude. It is therefore not orbiting within the plane defined by the DoS. However, its orbital pole is close to the MW disc pole (about $20^\circ$\ away).

\textbf{Acheron}: This stellar streams has most likely a globular cluster origin. A start and an end-position with distance-estimates are given in the discovery paper \citep{Grillmair2009}, which are used as anchor points. As in the cases of Cocytos and Lethe, the positions are uncertain by $0.5^\circ$. The distances for both stream positions are also reported in \citet{Grillmair2009}, their uncertainties are adopted to be 20 per cent, as discussed in our Section on the Lethe stream.

The stream-normal of Acheron is, with almost $79^{\circ}$, the second-furthest away from the DoS-normal.

\textbf{Cetus Polar Stream}: The anchor points for the stellar Cetus Polar Stream (CPS), are taken from the listed stream detections in table 1 (first and last entry) of the discovery paper by \citet*{Newberg2009}. The authors speculate that the CPS might be related to the globular cluster NGC 5824, which is classified as an OH object.

The stream-normal is the most distant one from the DoS normal, at an angular distance of $82^{\circ}$, which makes it almost perpendicular to the disc of satellites.

\textbf{Virgo Stellar Stream (VSS)}: This stellar stream or over-density is a problematic case and has not been included in the analysis as its direction is badly defined and its angular extend is rather short. The detection of the stream was reported by \citet{Duffau2006}. In that work, the stream's projection on the sky is aligned close to the celestial equator. 
However, \citet{Prior2009} find four additional RR Lyrae stars to be likely members of the stream, which extends the VSS further south but does not lead to a conclusive position angle on the sky for the stream. More light might be shed on this issue by the orbit-estimate for a single RR Lyrae Star identified as a VSS member \citep{Casetti-Dinescu2009}, which, for reasons of consistency of the two-anchor-point method, was not investigated in the present work.

\subsection{Results}
\label{sect:results}

In Fig. \ref{fig:normalvectorsbig} the positions of the resulting stream-normal vectors for the analysed streams are plotted. The plot includes the direction of the normal to the DoS derived from 24 satellite galaxies of the MW by \citet{Kroupa2010}. In addition, the normal vectors determined individually for the 11 'classical' and for the 13 'new', faint MW satellite galaxies are included \citep{Kroupa2010} (filled magenta squares with vertical and horizontal line, respectively). Also included are the normal vectors to the best fitting discs of globular clusters categorized as YH, OH and BD clusters (blue, yellow and cyan open diamonds, respectively) and the orbital poles of the MW satellites \citep{Metz2008} (green dots).
Even though the normal vectors are two-directional since we do not know the orbital sense, only one direction is plotted to avoid a too crowded plot. Only normals within the Galactic longitude range of $120^\circ < l < 300^\circ$\ are plotted. This range was chosen so that the Sagittarius stream-normal fits with the satellite's orbital pole while the majority of the stream-normals fall close to the average orbital pole of the MW satellite galaxies at ($l$, $b$) = ($177^{\circ}$,$-9^{\circ}$) (dark green diamond).

A clustering of stream-normal vectors close to $l = 160^{\circ}$\ is apparent, in the same direction to which the normal to the DoS points. Seven streams, 50 per cent of the streams in the analysis, are found within $32^\circ$\ of the normal vector to the DoS from \citet{Kroupa2010}. These are, with increasing angular distance, the GCN, NGC 5053, Styx, Magellanic, Cocytos, Palomar 5 and Lethe streams. Thus, all these streams most likely orbit within the plane of the DoS. The anchor points of the streams aligned with the DoS have Galactocentric distances as low as 11 kpc. This indicates that there is a vast polar structure (VPOS) consisting of satellite galaxies, streams and GCs, which might have an impact on the substructure as close as about 10 kpc from the MW centre. 
The angular distance $\Theta$\ of the stream-normal vectors does not correlate with the average Galactocentric distance of the stream anchor points.

Spherically averaging the angular coordinates of the stream-normal directions of all streams results in an \textit{average stream-normal} pointing to $l = 155^{\circ}$\ and $b = -26^{\circ}$. The spherical standard deviation \citep{Metz2007} is $\sigma_{\rm{sph}} = 46^{\circ}$. 
This is in close proximity ($24^{\circ}$\ away) to the normal vector of the DoS and is also close (about $27^{\circ}$\ away) to the average orbital angular momentum of the MW satellites \citep{Metz2008}. For ease of comparison, the average satellite orbital pole, available DoS normal fits and the average stream-normal directions found in this work are compiled in Table \ref{tab:normalpositions}. It is stunning to see how well they all agree with each other, given the different methods and objects used.
In addition, the normal vector of the plane fitted to the YH clusters, the DoGC$_{\rm{YH}}$, falls into the same region, while those of the OH and BD clusters are clearly distinct. 

Two of the four streams closest to the DoS most probably have their origin in dwarf galaxies: Styx and GCN. Another one, the Magellanic Stream, is already known to emanate from the Magellanic Clouds which are known to orbit within the DoS.
But there are also streams certainly or probably originating from dwarf galaxies with stream-normal directions further away from the DoS normal: the Orphan, Sagittarius, Anticentre and maybe the Cetus Polar Stream.

Four of the seven streams close to the VPOS have their origin in globular clusters.
Two of these streams are associated with known GCs of the MW, NGC 5053 and Palomar 5. Interestingly, both cluster are classified as YH objects, whose distribution shows a best-fitting normal vector close to the DoS normal (Sect. \ref{sect:clusters}). This is an intriguing sign of consistency. This situation resembles the finding for the MW satellites, whose positions define the DoS. Those satellite galaxies with known proper motions show orbital poles which also align with it \citep{Metz2008}.
Note that the outer clusters dominate the disc-fitting, while the two clusters with a stream aligning with the DoS are, with 16.9 and 18.6 kpc, at intermediate Galactocentric radii for YH objects. One third of the YH clusters have Galactocentric radii larger than 25 kpc. While the positions of the inner clusters are not crucial for the DoGC$_{\rm{YH}}$, their orbits might show signs of their accretion origin. This agrees with the finding that fitting the YH clusters with Galactocentric distances of less than 20 kpc individually also leads to a normal vector closely aligned with the DoS normal and thus is another sign that the VPOS reaches deep into the MW. 

However, the other stream associated with a YH cluster, NGC 5466 which lies at a slightly smaller Galactocentric distance of 16.2 kpc, has a stream-normal vector not aligned with the DoS or VPOS. It is reasonably close to the Sagittarius stream-normal and orbital pole, as well at the normal vector to the OH globular clusters. However, \citet{Law2010} conclude that NGC 5466 could not have been associated with Sagittarius, contradicting earlier findings by both \citet*{Palma2002} and \citet*{Bellazzini2003}. Scattering events might have changed any of the orbits though.

The Cetus Polar Stream, with a stream-normal almost perpendicular to the DoS normal, might be associated with the OH cluster NGC 5824, which \citet{Law2010} exclude as a former Sagitarrius member.

\subsection{Probability of the stream distribution}
\label{sect:probability}

The clustering of stream-normals close to the \emph{a-priori} and \emph{independently defined} DoS is intriguing. Could this be a chance result? 

A Kolmogorov--Smirnov test \citep[K-S test, e.g.][]{NumRec1992} evaluates how likely it is that a sample is drawn from a given distribution function. In the present case, this null-hypothesis is that the streams are drawn from an isotropic distribution function. 
The angular distances, $\Theta$, for the stream-normals from the DoS-normal have been calculated for each stream. Assuming an isotropic origin of the stream progenitors, the probability to find the observed distribution of these distances to be a random clustering can be estimated.

\begin{figure}
 \centering
 \includegraphics[width=80mm]{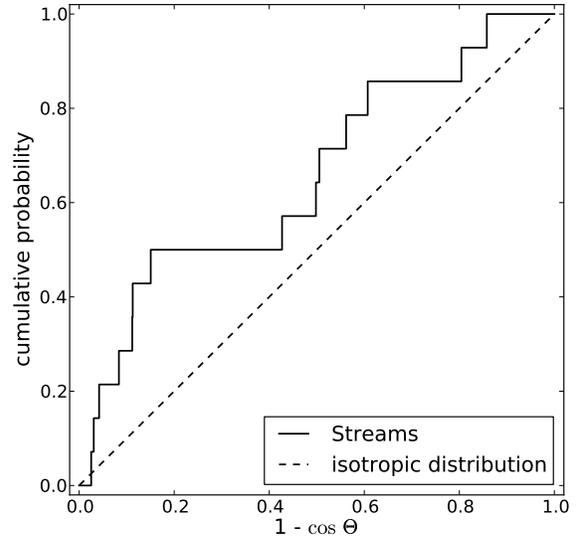}
 \caption{Cumulative probability of stream-normals within a fraction of the volume on the half-sphere measured from the DoS normal vector. Plotted are the observed stream-normal vectors (solid line) and the expectation for an isotropic distribution (dashed line). The two distributions clearly differ as there are 7 out of 14 streams found within the inner 16 per cent of the available area.}
 \label{fig:angledist}
\end{figure}

The cumulative distribution of angular distances $\Theta$\ of stream normals to the DoS normal are plotted in Fig. \ref{fig:angledist}. The angles are expressed by which fraction of the area of a half-sphere is enclosed within a circle of the same radius as the angular distance $\Theta$\ to the stream. In an isotropic distribution the probability for a stream-normal's direction is proportional to the area, so an isotropic distribution is represented by a straight, diagonal line with a slope of one in Fig. \ref{fig:angledist}. The plot makes the reported clustering of stream-normals around the DoS normal obvious: 7 out of 14 streams are centred within less than $1/6^{\rm{th}}$ of the available area.

For the K-S test, the maximum distance $D$\ between the cumulative probability of the stream-sample and of the isotropic case is calculated. In the K-S test $D$\ is a measure for the likelihood of the sample being drawn from the distribution function. For the analysed sample of 14 streams, $D$ is found to be 0.349. Using the approximate formula in \citet{NumRec1992}, the K-S test can thus rule out the null-hypothesis with 95 per cent (about $2 \sigma$) confidence.

However, the K-S test does not include the full knowledge about the clustering of the stream-normals close to the DoS. In fact, a similar $D$-value could be produced if half of the streams would lie far from the DoS normal, but at similar angular distances. The K-S test does not give the probability to find 7 out of 14 streams \textit{close to} the DoS, but to find them \textit{at similar angular distances} from the DoS normal. The test does not use the full information available, represented by the direction of the DoS normal vector.
Therefore, a second, independent method to estimate the probability that the observed distribution could be explained as a chance-result of a uniform distribution is asked for.

It can be calculated how probable it is that at least $k$\ out of $n$\ stream-normals are found within an angular distance $\Theta$\ to the DoS normal. The alignment of the stream-normal over-density with the DoS-normal has to be factored in. Without that, the probability that $k$\ uniformly distributed streams lie close to each other would be higher, but the absolute position of this clustering would be undefined.

The probability $p(\Theta)$\ of one vector randomly chosen on the sphere to point into the area $\Theta$\ degrees around the independently defined DoS-normal is given by the area within the circle of radius $\Theta$\ degrees on the sphere, $A(\Theta)$, in relation to the total area of the half-sphere\footnote{Because the normal vectors defining the same plane can point in two opposing directions, so only one half-sphere has to be considered.}, $A(90^\circ)$:
\begin{displaymath}
p(\Theta) = \frac{A(\Theta)}{A(90^\circ)} = \frac{1 - \cos \Theta}{1}.
\end{displaymath}
This has to be modified because the assumed isotropy of the normal distribution can not hold any more with stream-normals of high Galactic latitudes, as these are close to the MW poles. Streams with normals having high $\vert b \vert$-components will orbit within the MW disc, these are thus on the one hand hard to detect when far from the Sun, which will result in completeness issues, and on the other hand their origin could also be related to the MW disc dynamics and its build-up. We therefore exclude these regions around the Galactic poles, defined by the angle $\beta$, from the calculation of the probability, which reduces the effective total area of the half-sphere. The probability for one normal vector then is given by 
\begin{displaymath}
p(\Theta, \beta) = \frac{1 - \cos \Theta}{\cos \beta}.
\end{displaymath}
Note that this is only valid if the two regions around the DoS normal and the MW pole do not overlap, so if the sum of $\Theta$, $\beta$\ and the distance from the DoS normal to the equator ($2.2^\circ$) is at most $90^\circ$.

The probability, given by a common Bernoulli experiment, for one specific set of $k$\ streams out of $n$\ being within $\Theta$\ degrees to the DoS is
\begin{displaymath}
p(\Theta, \beta)^{k} \left[ 1 - p(\Theta, \beta) \right] ^{n-k}.
\end{displaymath}
There are, however,
\begin{displaymath}
{n \choose k}  = \frac{n!}{(n-k)! \cdot k!}
\end{displaymath}
possibilities to draw $k$\ streams out of a total of $n$, allowing each stream to be drawn once at most independent of the order. Thus, the probability to find exactly $k$\ streams out of $n$\ within $\Theta$\ degrees of the DoS is
\begin{displaymath}
{n \choose k} p(\Theta, \beta)^{k} \left[ 1 - p(\Theta, \beta) \right] ^{n-k}.
\end{displaymath}
The probability to find \textit{at least} $k$\ streams out of $n$\ within this area is then given by the sum over the probabilities to find $k, k+1, ..., n$\ streams within the area:
\begin{displaymath}
P = \sum\limits_{i=0}^{n-k} {n \choose k+i} p(\Theta, \beta)^{k+i} \left[ 1 - p(\Theta, \beta) \right] ^{n-k-i}.
\end{displaymath}

An exclusion region of $\beta = 20^{\circ}$\ around the MW pole makes the Anticentre stream-normal to lie barely within the allowed region but excludes streams orbiting within the disc of the MW galaxy. 7 stream-normals within $\Theta = 32^{\circ}$\ of the DoS are found, which has a probability of $P(\Theta=32^{\circ},\beta=20^{\circ},n=14,k=7) = 0.34~\rm{per cent}$\ if the streams were on isotropically distributed orbits. 
Not considering the exclusion region gives a slightly smaller probability of $P(\Theta=32^{\circ},\beta=0^{\circ},n=13,k=6) = 0.24$\ per cent.

\subsection{Anchor point distance effects}
\label{sect:anchorpointdistanceeffects}

\begin{figure}
 \centering
 \includegraphics[width=80mm]{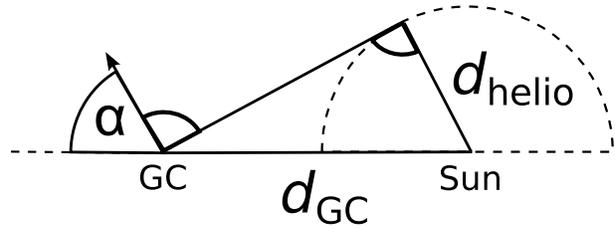}
 \caption{Sketch illustrating the minimum angle $\alpha$\ a stream-normal vector can have from the axis connecting the Sun and the Galactic centre (GC in the sketch), which lies at Galactic coordinates $(l,b) = (0^\circ,0^\circ)$. The Heliocentric distance of the stream anchor point is $d_{\rm{helio}}$\ and the distance from the Sun to the Galactic centre is $d_{\rm{GC}}$. In this estimation, the angle $\alpha$\ is minimal when the line connecting the Galactic centre is a tangent to the circle of radius $d_{\rm{helio}}$\ around the Sun. Thus, a stream with anchor points closer than $d_{\rm{helio}}$\ can never have a stream normal vector closer than $\alpha$\ to $(l,b) = (0^\circ,0^\circ)$\ and $(180^\circ,0^\circ)$.}
 \label{fig:alphasketch}
\end{figure}

Fig. \ref{fig:alphasketch} illustrates that a stream defined by anchor points with Heliocentric distance $d_{\rm{helio}}$\ smaller than the distance between the Sun and the Galactic centre $d_{\rm{GC}}$\ can not lead to stream-normals closer to $(l,b) = (180^\circ,0^\circ)$\ or $(0^\circ,0^\circ)$ than
\begin{displaymath}
\alpha = \arctan \sqrt{ \left(\frac{d_{\rm{GC}}}{d_{\rm{helio}}}\right)^2 - 1}.
\end{displaymath}
The DoS normal is close to $(l,b) = (180^\circ,0^\circ)$. Therefore, when streams close to the Sun can be assumed to have a higher chance of detection, \textit{streams not aligned with the DoS might be over-represented in the sample}.

The position of us as observers thus might prefer the finding of stream positions that are outside of the DoS. But still, the stream-normal overdensity is strong enough to be visible in the data analysed. The previous probability-estimates have thus to be considered as upper limits only. When excluding streams with $d_{\rm{helio}}$\ that have no chance of giving normals close to DoS (only Acheron in this sample), the probability for the streams to be aligned in the way they are when drawn from a random distribution is further reduced to $P(\Theta=32^{\circ},\beta=20^{\circ},n=13,k=7) = 0.20$\ per cent.

\section{Discussion}
\label{sect:discussion}

\begin{figure}
 \centering
 \includegraphics[width=80mm]{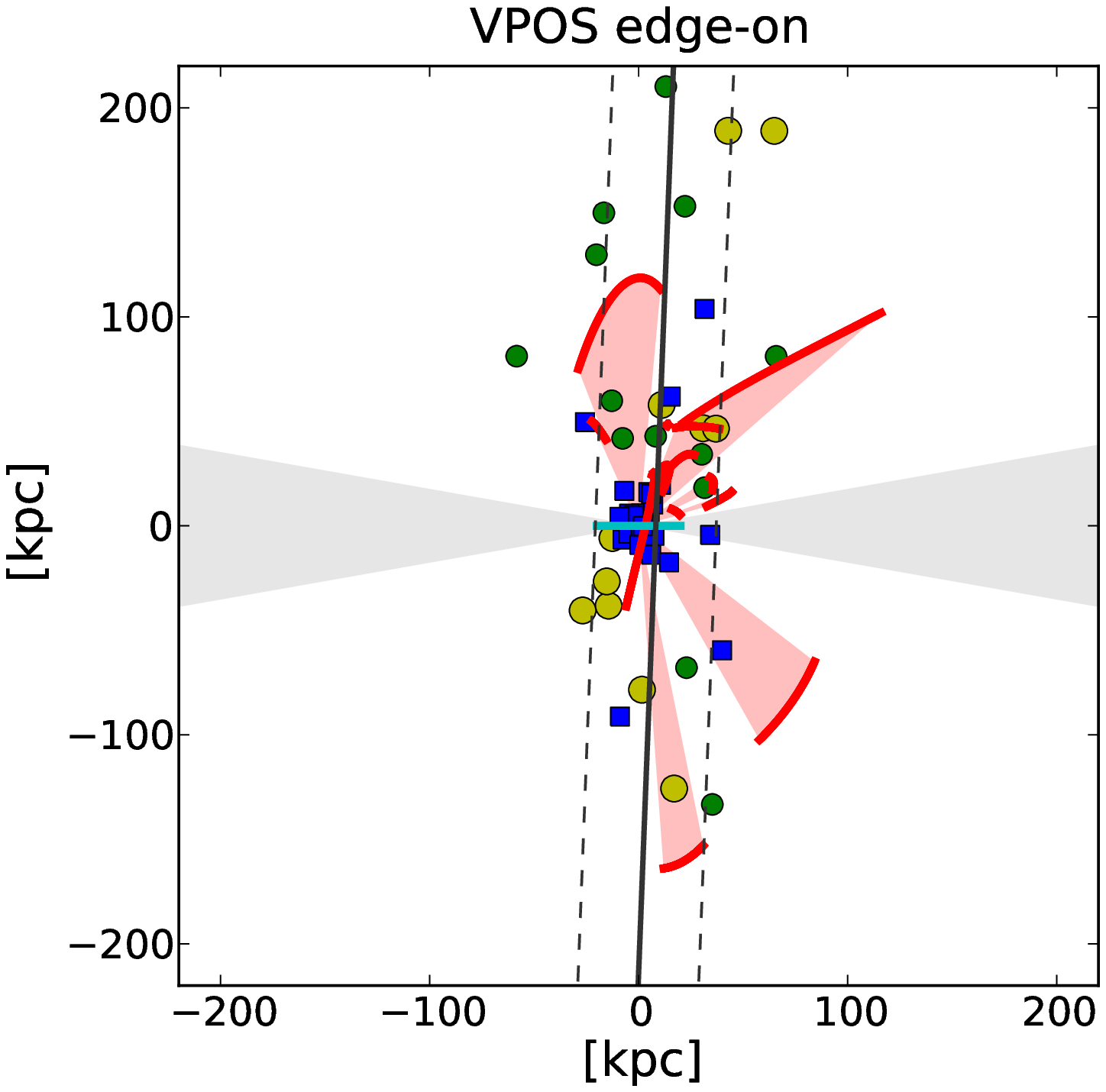}
 \includegraphics[width=80mm]{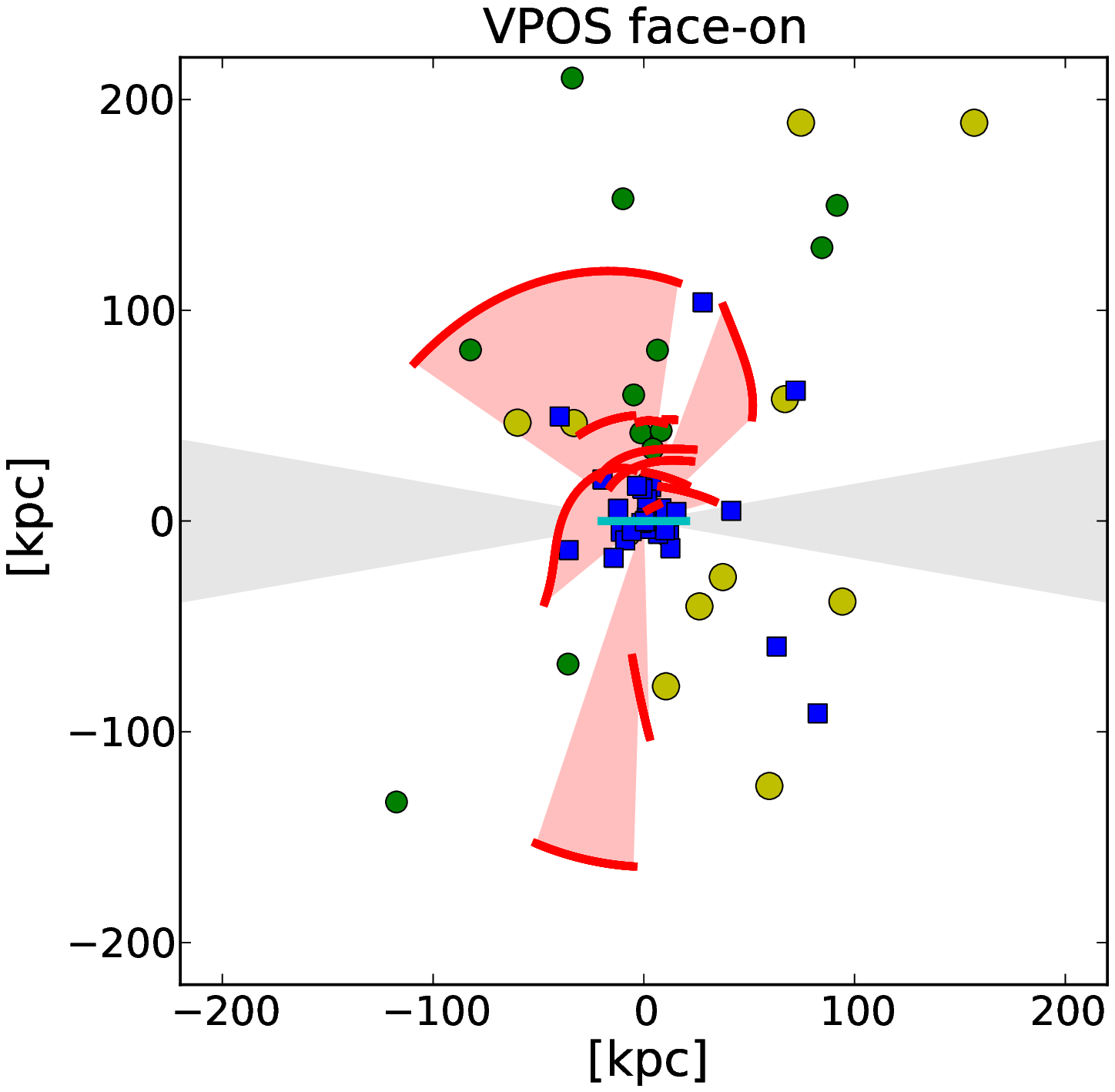}
  \caption{The vast polar structure -- VPOS -- about the MW in Cartesian coordinates. The y-axis in both panels points towards the Galactic north pole. The 11 classical satellites are shown as large (yellow) dots, the 13 new satellites are represented by the smaller (green) dots, YH GCs are plotted as (blue) squares. The (red) curves connect the stream anchor points, the (light-red) shaded regions illustrating the planes defined by these and the Galactic centre. Note that the stream coordinates are magnified by a factor of 3 to ease the comparison. The obscuration-region of $\pm 10^\circ$\ from the MW disc is  given by the horizontal grey areas. In the centre, the MW disc orientation (edge-on) is shown by a short horizontal (cyan) line. \textit{Upper panel}: an edge-on view of the VPOS (x-axis aligned with $l = 156.4^\circ$). The near-vertical solid line shows the best fit Disc of Satellite galaxies -- DoS -- (seen edge-on) at the given projection from \citet{Kroupa2010}, the dashed lines define the RMS-height of the DoS (28.9 kpc). Both YH GCs and streams align in a similar direction. \textit{Lower panel}: a view rotated  by $90^\circ$, the VPOS is seen face-on. The satellite galaxies and YH GCs show a wider distribution and the planes of most streams are also seen close to face-on. A movie rotating this view over $360^\circ$ can be found online at http://www.astro.uni-bonn.de/download/.}
 \label{fig:DoSfaceedge}
\end{figure}

In this section we present our interpretation of the results and discuss possible explanations and implications. 

Using a new method to estimate the normal to the orbital plane a stream forms around the Galactic centre, a pronounced overdensity of stream-normal vectors is unveiled. It is stunning that half of the streams are well aligned (having normal vectors closer than $32^\circ$, the next-closest one being $55^\circ$\ away) with the previously known DoS, with the average orbital angular momentum of six nearby satellite galaxies, and with the normal vector to the best-fitting plane through the YH globular clusters. This is especially noteworthy, as the streams analysed not only include debris of tidally disrupted dwarf galaxies, but also consist of the remnants of globular clusters. The possibility of a similar alignment to occur by chance if the stream orbits are drawn from an isotropic distribution is estimated to be 0.3 per cent. 

The radial distances from the MW centre of the considered objects show that the polar structure reaches deep into the MW, with stream anchor points having Galactocentric distances of 10-50 kpc, the YH GCs aligned close to the DoS having distances between 20 and 100 kpc and satellite galaxies found close to the DoS ranging as far out as 250 kpc. As it consists of different types of objects, the structure is termed a vast polar structure (VPOS) around the MW, to distinguish it from the disc of satellites (DoS) which only consists of the satellite dwarf galaxies of the MW. The three-dimensional distribution of the objects in the VPOS is illustrated in Fig. \ref{fig:DoSfaceedge}, in which the streams are magnified by a factor of three.

\subsection{Might the MW be an ancient polar-ring galaxy?}

The VPOS shows that the MW has a pronounced planar, polar structure extending from 10 kpc to its outermost region of 250 kpc. It can therefore be termed a polar-disc galaxy, a term which has also been suggested as a more correct replacement to the term polar-ring galaxy \citep[e.g.][]{Gallagher2002, Brook2008}. Could the MW have been a polar-ring/disc galaxy?

The MW shares many properties with the criteria defined by \citet{Whitmore1990} for the classification as a polar-ring-type galaxy.
The major axes of the two components, the MW disc and the VPOS, are nearly orthogonal, as are their angular momenta, shown by the satellite galaxy orbital poles \citep{Metz2008} and the stream normal vectors clustering close to the DoS normal. 
The centres of the MW and the VPOS are nearly aligned, illustrated for example by the small offset of only 8 kpc of the DoS from the Galactic centre \citep{Kroupa2010} and 2.6 kpc for the best-fitting plane to the YH GCs. Finally, the VPOS consists of luminous material and is thin as the half-thickness or height of the DoS is 28.9 kpc, while the radius is $\approx 250$\ kpc. Polar rings formed in numerically modelled galaxy mergers and interactions are stable for long times and can be transformed into polar discs when the ring is disturbed by other galaxies, such as the later merger of the ring's donor galaxy in an interaction \citep{Bournaud2003}.

Speaking against the classification as a polar-ring/disc galaxy are the size of the VPOS (more extended than typical polar rings/discs) and the fact that the VPOS consists dominantly of individual stellar subsystems and not of more evenly spread-out material. However, until the long-term evolution of polar-ring galaxies is better understood to allow a decisive answer, the MW might be interpreted as a candidate for an \textit{ancient remnant} polar-ring galaxy.

\subsection{The polar structure in the scenario of satellite accretion}

The pronounced VPOS can not be expected in the accretion scenario of individually infalling satellite galaxies. The YH clusters are popularly thought to be accreted objects too, which might have been brought in along with dwarf galaxies, explaining the similar imprint in their spatial distribution. Both types of accreted objects will be tidally disrupted, giving rise to streams in the MW halo. Naturally, there will be a preference to orbit in a similar direction, indicated by the stream-normal vectors aligning with the DoS and DoGC$_{\rm{YH}}$. 

The GCs, after having been stripped off their dwarf galaxy, would have orbits similar to that of their former host. Within the cosmological scenario, the distribution of YH GCs thus adds information about the preferred orbit of their hosts, even if those might not be around any more. Consequently, hosts not orbiting within the VPOS will produce GCs not being distributed within this structure. If the VPOS is a young structure formed by group infall of DM subhalos, most YH GCs must have been stripped recently from dwarf galaxies belonging to that group. Then, however, the YH GCs stripped longer ago from galaxies with unrelated orbits are missing. 
If the satellites fell in along dark-matter filaments, as suggested by \citet{Lovell2011}, the majority of subhalos (65 to 78 per cent or more) still would not be distributed in the preferred plane, as their work has shown. As a consequence, a similar fraction of the dwarf galaxies populating these subhalos and therefore their GCs are predicted to be located and orbiting outside such a plane.

In order to account for the observed large-scale alignment of various subsystems within the standard cosmological model, individual dwarf galaxies with their GCs would have to fall into the MW dark matter halo in virtually the same plane highly inclined to the MW disc. However, there is no consistent explanation for a preference to accrete cosmological objects in a plane perpendicular to the MW (see the discussion in Sect. \ref{sect:introduction}).

\subsection{The MW satellites as TDGs}
\label{sect:MWsatsasTDGs}

\begin{figure*}
 \centering
 \includegraphics[width=170mm]{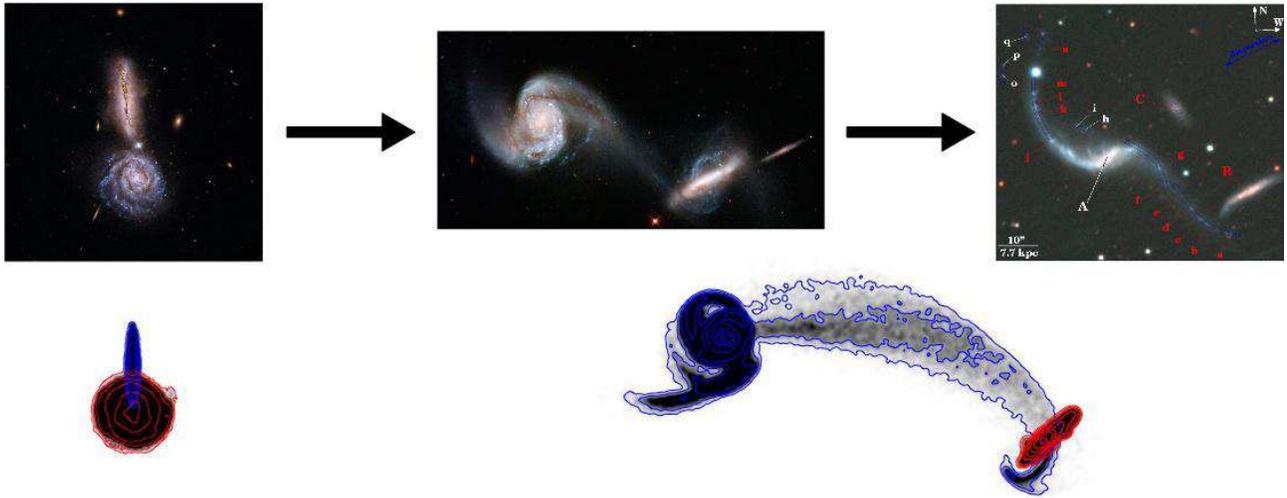}
 \caption{ Three examples of (near-) polar interactions observed at different stages. They illustrate qualitatively the possible interaction of the MW that resulted in the formation of its satellite galaxies as TDGs. Two approximately perpendicular disc galaxies (left picture, Arp 302) collide on a polar orbit. Material gets tidally stripped and is accreted in a polar structure around the other galaxy (central picture, Arp 87, which \citet{Whitmore1990} classify as an object possibly related to polar-ring galaxies). In the tidal tails dwarf galaxies form (right picture, the Dentist's Chair galaxy), where the galaxies labelled A and B are the interaction partners and knots along the tidal tails labelled with lower-case letters are TDG candidates. They will orbit one of the two interacting galaxies in a plane defined by the interaction geometry. The interaction can end in a merger or in a fly-by of the two major galaxies.
 The lower row shows the particle densities of two snapshots from a model of two galaxies interacting in a similar geometry \citep[model 5deg200vel from ][]{Pawlowski2011}. They were generated by smoothing individual particle positions with a Gaussian and then plotting contours of constant density. Before the interaction (left) the target galaxy (red contours) is seen face-on along the plane of the interaction. The second snapshot (right) shows a view projected onto the plane of the interaction, with the target galaxy seen edge-on to the right. The orbital poles of model particles of the infalling galaxy (blue contours) at the end of this calculation are plotted in the lower panel of Figure \ref{fig:modelpoles}.
Image Credits:  
Arp 302: NASA/STScI/NRAO/ A. Evans et al.;
Arp 87: NASA, ESA, and the Hubble Heritage Team (STScI/AURA); the 'Dentist's Chair' Galaxy AM 1353-272: \citet{Weilbacher2002}.}
 \label{fig:pictures}
\end{figure*}

In contrast to the cosmological scenario, a tidal scenario for the formation of the MW satellites as tidal dwarf galaxies \citep[TDGs, for an introductory review see][]{Bournaud2010} naturally resolves the seemingly problematic correlation of satellite galaxies in phase-space \citep{Pawlowski2011}. In case of such a tidal origin of the MW DoS, large numbers of stellar clusters form together with the TDGs, as \citet{Bournaud2008} have shown in a high-resolution simulation of a galaxy merger. This explains the similarity of the phase-space distribution of satellite galaxies, YH GCs and streams ranging from 250 kpc down to 10 kpc Galactocentric distance in the VPOS.

The polar structure surrounding the MW is ancient. Its age can be estimated from the age of its constituents. \citet{Mackey2004} have discussed the ages of the different populations of GCs. They used relative ages from \citet{Salaris2002}, which are given with respect to the age of the cluster M92. Assuming an age of 12.55 Gyr for the cluster M92 \citep[like in][] {Mackey2004}, the ages of the 30 OH GCs with age estimates show a strong peak at about 12.2 Gyr. The 17 YH clusters with age estimates show two peaks, one close to the one of the OH population at 11.8 Gyr, and a younger one at 9.3 Gyr. If the formation of the VPOS has been triggered by a major event, this therefore must have happened 9 to 12 Gyr ago.

We suggest that the MW has experienced a near-polar collision with an approximately perpendicularly oriented disc galaxy. Figure \ref{fig:pictures} illustrates this geometry with images of observed interacting galaxies and two model snapshots. They show that polar interactions of perpendicularly oriented disc galaxies happen even at the current epoch and result in tidal debris distributed in a polar structure.

The initial interaction could have been a fly-by of two galaxies or it might have ended in a galaxy-merger, destroying the infalling galaxy in the process. In a major merger, the two colliding galaxies will form a spheroidal object. As this event must have happened about 10 Gyr ago, the MW disc might have re-formed from gas accreted later \citep{Hammer2005}, with the spheroidal component being the bulge of the MW today. \citet{Ballero2007} estimate that the MW bulge must have formed rapidly on a time scale of 0.1 Gyr, in favour of a merger-induced origin. However, also in a fly-by encounter material stripped from the passing galaxy would be accreted onto the MW, possibly producing a bulge component. A bar instability would in any case channel gas onto radial orbits.

If the TDGs were not formed in a merger but in a fly-by encounter, the passing galaxy still has to be nearby. Two candidates can be found in the Local Group: Andromeda and the LMC.

The Andromeda galaxy M31 is a promising candidate because it is the biggest nearby galaxy and thus a large amount of matter could have been stripped from it to form TDGs. Furthermore, the distribution of dwarf galaxies around M31 is anisotropic, too \citep{Metz2007}, with a disc of satellites fitted to the M31 dwarfs oriented such that it is seen nearly edge-on from the MW \citep{Metz2009}. This can also be seen from the positions of M31 dwarf galaxies on the sky plotted in figure 1 of \citet{Tollerud2011}. The M31 satellites are preferentially distributed in a structure extending approximately from north to south in Galactic coordinates, just as the MW VPOS extends in north-south direction. A common direction of the satellite distributions of both galaxies is expected in a tidal scenario that formed both satellite populations together, as TDGs form in a plane defined by the orbit of the interaction.

The scenario in which the satellite galaxies have been formed as TDGs in a past passage of the MW and M31 therefore puts constraints on the orbits of those two galaxies, making it possible to predict the direction of proper motion for M31. This was done by \citet{Sawa2005}, who suggest that the dwarf galaxies of the Local Group were formed in an off-centre collision of the MW and M31 some 10 Gyr ago. While they do not consider TDG formation but dwarf galaxy formation due to compression of halo gas during the closest approach of the two galaxies, the orbital constraints are similar. As \citet{Zhao2010} estimate from a timing argument, even in modified Newtonian dynamics (MOND) M31 and the MW might have had a fly-by encounter in the past. Furthermore, as dynamical friction is much weaker in MOND (due to the absence of dark matter), galaxy-galaxy interactions ending in a merger are less frequent, while fly-bys can be considered more common MOND \citep{Tiret2008}.

Another hint towards this scenario might be the fact that M31 shows a number of features which might be explained by a violent interaction with another galaxy in the past. For example, \citet{Hammer2010} have shown that a number of features -- the morphologies and kinematical properties of the thin and thick disc, bulge and streams in the halo of M31 -- can be explained by a single major merger about 9 Gyr ago. They also mention the formation of TDGs in their simulations, which might contribute to M31's satellite galaxy population or could have been ejected into the direction of the MW \citep{Yang2010}. However, if only a few TDGs of M31 have been accreted by the MW, the probability to find them in a pre-existing, independently formed VPOS is low. Fouquet et al. (in prep.) therefore suggest that the 11 classical dwarf galaxies are TDGs that originate from a major merger occurring in the history of M31. They formed in the first-passage tidal tail and were expelled into the direction of the MW where they now form a part of the VPOS. On the other hand, a fly-by scenario between the young MW and the young Andromeda galaxy should also be investigated. Such an interaction with an approximately equal-mass galaxy might have had similarly strong effects on the M31 galaxy and enrich both galaxies with satellite galaxies formed as TDGs.

In the scenario first suggested by \citet{LyndenBell1976} and further discussed in \citet{Pawlowski2011}, the LMC progenitor galaxy might have been the origin of the MW satellite system. 
If the MW satellites have formed from material stripped from the LMC progenitor galaxy, the LMC needs to be on a bound orbit and at least on its second approach to the MW. This was questioned by the recent proper motion measurements of the LMC by \citet{Kallivayalil2006} and \citet{Piatek2008}. These revised proper motions could result in an unbound or wide orbit \citep{Besla2007, Wu2008, Ruzicka2010}, which had let \citet{Yang2010} to argue that the LMC might be a TDG expelled from a merger event in Andromeda. However, one has to keep in mind that the local standard of rest (LSR) circular velocity is probably also higher than previously assumed \citep{Reid2004, Reid2009}. As \citet{Ruzicka2010} have shown, an increase in the LSR circular velocity increases the total mass of the MW, thus increasing the total potential energy of the LMC with respect to the MW, while at the same time reducing the galactocentric velocity of the LMC and thus decreasing the kinetic energy. In total, a higher LSR circular velocity makes the LMC more bound again. Consequently, assuming a higher LSR velocity, the Magellanic clouds were shown to likely be bound to the MW \citep{Shattow2009, Ruzicka2010, Bartoskova2011}.

Another argument in favour of a bound orbit is the connection of the LMC to the VPOS: The LMC lies within the DoS, the Magellanic stream shows a stream normal vector close to the DoS normal and the orbital pole of the LMC is close to the DoS normal, having an angular distance of only $20^\circ$. This would have to be a coincidence if the LMC were unrelated to the VPOS and falling in for the first time. The probability for the LMC to show an orbital pole this close to the DoS normal and orbiting in the same direction as the majority of satellite galaxies with measured proper motions is only 3 per cent when assuming the probability distribution of its infall direction on the sky is uniform. Thus it is unlikely that the LMC orbit is unrelated to the VPOS.

It therefore seems to be possible that the LMC is on a wide but bound orbit and might have been the original collision partner of the MW which formed a population of TDGs. If the LMC is bound to the MW, this has another consequence.
The orbits of MW satellites (GCs or galaxies) might be changed by scattering events, such that the present orbits (and consequently stream-normal vectors) might not trace the initial configuration. In particular, \citet{Zhao1998} suggests Sagittarius to have had an encounter with the LMC and been scattered onto its present high-energy orbit, therewith solving the hitherto unsolved problem of its short orbital period. If this is true, then Sagittarius might have been a member of the VPOS until about 2-3 Gyr ago. If the LMC were on its first infall, a collision of the LMC and Sagittarius is unlikely or impossible. While the indications for a bound orbit of the LMC can be interpreted in favour of such an encounter, a decisive answer on the validity of the scattering scenario is currently not possible.

The gaseous stream GCN, having the closest stream normal to the DoS, might be an ancient remnant of the initial tidal tail from which the satellite galaxies and globular clusters formed. This interpretation is supported by its stream normal direction, which points close to the LMC orbital pole and is situated in the middle between the average orbital pole of the satellite galaxies, the average stream normal, the normal to the YH GCs and the DoS normal. It might thus indicate best the original direction of the colliding galaxy. Interestingly, \citet{Winkel2011} point out that a model calculation of \citet{Diaz2011} suggests that GCN might be associated with the leading arm of the Magellanic system.

\subsection{Questioning the TDG-scenario}

If the MW satellites are TDGs, then the following questions arise. For some, definite answers can not be given as no detailed studies aiming at this possible formation scenario for the MW satellites have been performed yet.

\subsubsection{Do TDGs align and orbit in a disc?}
\label{sect:tdgsindisc}

\begin{figure}
 \centering
 \includegraphics[width=85mm]{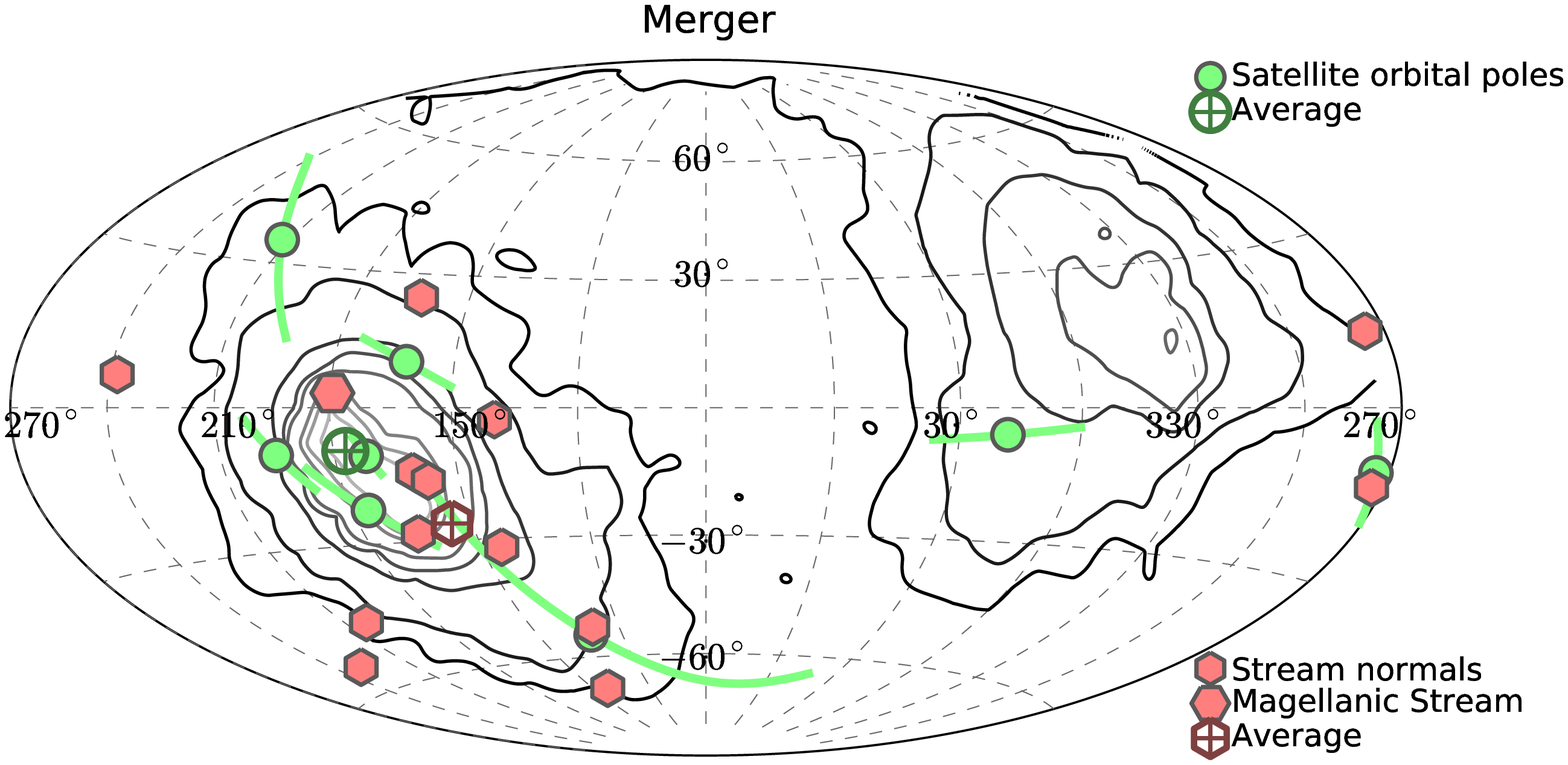}
 \includegraphics[width=85mm]{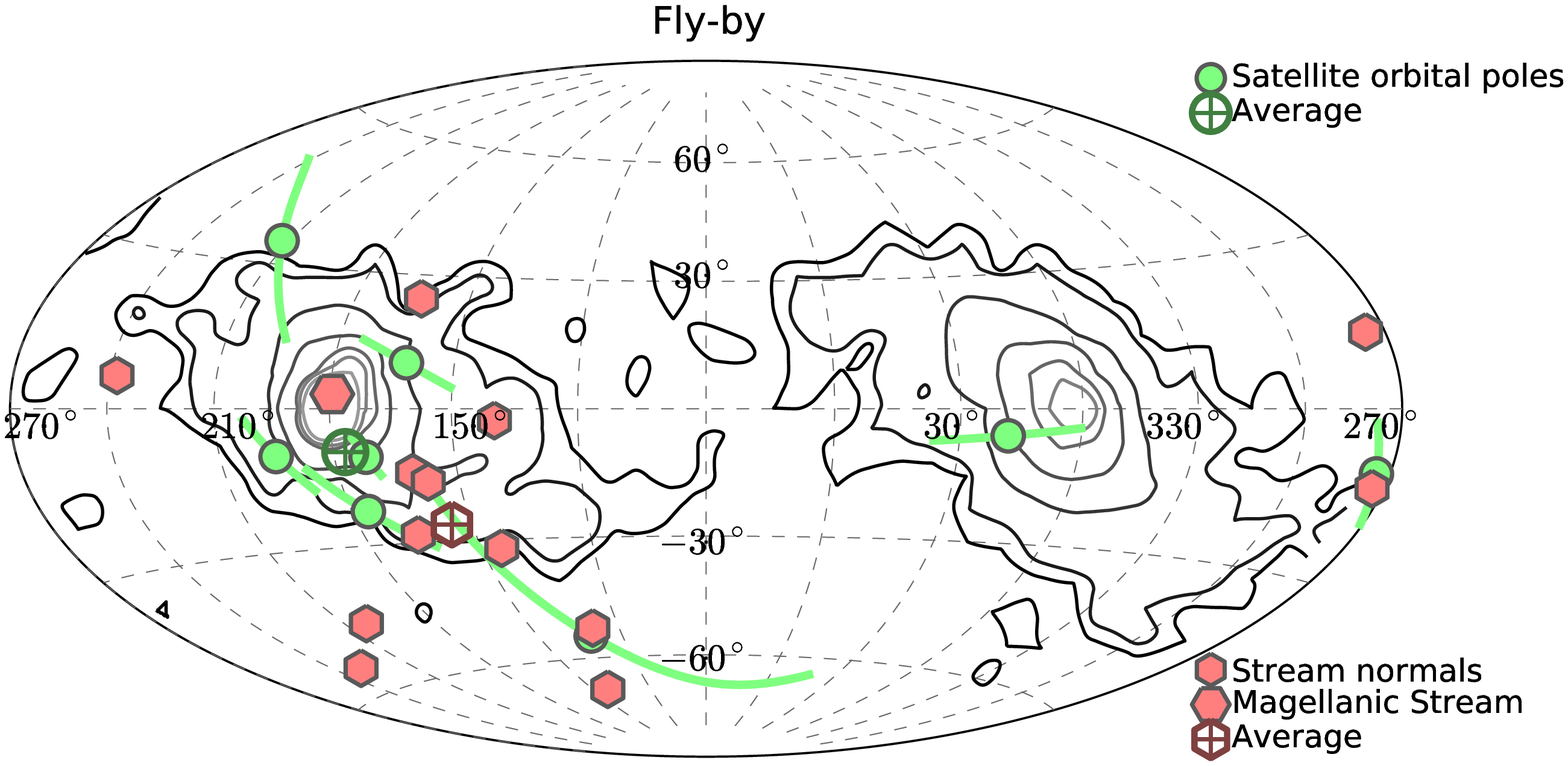}
 \caption{The directions of the orbital poles of the MW satellite galaxies from \citet{Metz2008} and the stream normal directions (this work) compared to the distributions of orbital poles in tidal debris of galaxy-interaction models \citep{Pawlowski2011}, plotted as contours. Note that the $l$-axis is shifted by $90^\circ$ compared to Fig. \ref{fig:normalvectorsbig}. The \textit{upper panel} shows the result of a merger calculation, the \textit{lower panel} those of a fly-by. The orbital poles of all particles in the models that have distances of at least 20 kpc in the final snapshot are included. The coordinate system of the models is defined in analogy to the Galactic coordinates, with the spin direction of the target galaxy pointing to $b = -90^\circ$\ in the merger and to $b = 90^\circ$\ in the fly-by model. The initial orbital pole of the infalling galaxy in both models points to ($l$, $b$) = ($180^\circ$, $0^\circ$). The contours show the densest regions including 95, 90, 80, 70, ... per cent of all particles. The agreement of the observed satellite orbital poles and stream normal directions with the contours of prograde material (left, centred on $l = 180^\circ$) is striking, but also retrograde material close to the orbital pole of Sculptor is found. Even the diagonal elongation of the orbital poles and stream normals (from the upper left to the lower right) is reproduced.}
 \label{fig:modelpoles}
\end{figure}

\begin{figure}
 \centering
 \includegraphics[width=80mm]{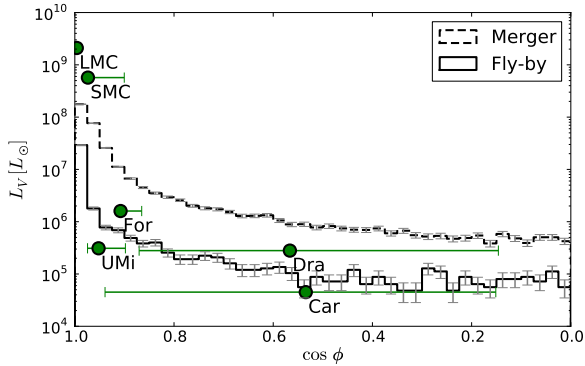}
 \caption{The luminosities of MW satellites co-orbiting in the VPOS. The y-axis shows the V-band luminosities $L_{V}$ of the MW satellites in units of $L_{\sun}$\ as compiled in \citet{Kroupa2010}. The x-axis plots the cosine of the angular distance $\phi$\ of the respective satellites' orbital pole from the average direction of the six satellites' orbital poles. 
 For comparison, the histograms show the distribution of orbital poles in tidal debris in two models. They result from the same data as Fig. \ref{fig:modelpoles}, assuming each particle to have 8,000 $L_{\sun}$, corresponding to a mass-to-light ratio of 2 for the model unit scaling in \citet{Pawlowski2011}. Here, $\phi$\ is measured from the median direction of the particles' orbital poles. The grey error bars show the uncertainties for each bin from Poisson statistics.
 The satellites with orbital poles further away from the mean direction (see Table \ref{tab:normalpositions}) are less luminous. In the TDG scenario this might indicate that less material for their formation was available on these orbits, which seems to be qualitatively confirmed by the histograms. The amount of tidal material with orbital poles within a certain distance-bin follows a similar trend as the MW satellite luminosities. No such trend is expected if the satellite galaxies would have fallen onto the MW on individual orbits.}
 \label{fig:lumcosphi}
\end{figure}

In contrast to the cosmological attempts to explain the VPOS, it can be considered as observationally confirmed that TDGs show planar distributions. As they form in tidal tails, observed TDGs show a strong spatial alignment with each other. Examples are the supermassive star clusters forming in the tidal tail of the Tadpole galaxy \citep{Jarrett2006} and the numerous TDG candidates that have been observed aligned along the tidal tails of interacting galaxies like NGC 5291 \citep{Malphrus1997} and AM 1353-272 \citep{Weilbacher2002}. This spatial alignment can be conserved over time. \citet{Galianni2010} discuss that the satellite galaxies of NGC 1097, embedded in a tidal stream, might be old TDGs. 
Recently, \citet{Duc2011} found three old TDG candidates, aligned along a tidal tail around the galaxy NGC 5557. 

\citet{Pawlowski2011} have modelled collisions of galaxies in which a target disc galaxy, representing the early MW, interacts with a perpendicularly oriented disc galaxy falling in on a polar orbit (left plot in Figure \ref{fig:pictures}). The interactions lead to polar discs of tidal debris, particles representing the debris orbit within these disc and even counter-orbiting material is a natural occurrence, possibly explaining the counter-orbit of the Sculptor dwarf galaxy. For two models, a merger and a fly-by \citep[named model 7.5deg100vel and 5deg200vel respectively in][]{Pawlowski2011}, the distributions of orbital poles of particles at the end of the runs is plotted in Figure \ref{fig:modelpoles}. 

A striking agreement of the observed satellite orbital poles and stream normal directions with the contours of prograde material is apparent, even the diagonal elongation (from the upper left to the lower right) is reproduced. At lower contours, further away from the median orbital pole of the particles, the density of tidal material is lower. This might indicate that TDGs formed on such orbits might preferentially be less-luminous as less material for their formation is available. 

The observed orbital poles of the satellite galaxies seem to be in agreement with this expectation, as Figure \ref{fig:lumcosphi} illustrates. The satellites with orbital poles furthest from their average direction, Carina and Draco, are the least luminous \citep{Mateo1998}, followed by the brighter Ursa Minor and Fornax, which are closer in, and then the the most luminous SMC and LMC in the centre.

The phase-space distribution of TDGs, as expected from tidal debris, is therefore in good agreement with the observed distribution and motion of MW satellite galaxies in a VPOS.

\subsubsection{Do enough TDGs form in interactions?}

In many interacting galaxies young or currently forming star clusters and TDG candidates have been found, for example in the ''Mice'' galaxies NGC 4676 \citep{deGrijs2003}; the ''Antennae'' galaxies NGC 4038/39, NGC 3256 and NGC 3921 \citep{Knierman2003}; Arp 105, Arp 242, NGC 7252 
\citep{Bournaud2004}; the ''Tadpole'' galaxy Arp 188 \citep{Jarrett2006}; Arp 305 \citep{Hancock2009} and  NGC 5557 \citep{Duc2011}.
Numbers of TDG candidates in the order of 10 have been observed in the tidal tails of interacting galaxies, as in the ''Dentist's Chair'' galaxy AM 1353-272 \citep{Weilbacher2002} and in NGC 5291 \citep{Malphrus1997}. These are relatively massive object (typically $> 10^8 M_{\sun}$), equivalents of the dSph and ultra-faint dwarf galaxies (typically $< 10^7 M_{\sun}$) of the MW would have too low a surface brightness to be observed in distant galaxies.

In a hydrodynamical high-resolution modelling of a galaxy merger, \citet{Bournaud2008} find more than 100 new-formed objects consisting of young stars. These TDGs and smaller systems have masses in the range $10^5~\rm{to}~10^8~M_{\sun}$. The colliding galaxies of \citet{Bournaud2008} are rather gas-poor, with a gas fraction of only 17 per cent they represent a galaxy merger at $z \leq 1$. This is important as the numerical models of \citet{Wetzstein2007} have shown that the gas plays a major role in the formation of TDGs, such that an encounter in the early universe of galaxies consisting of more gas can be expected to form more or more-massive TDGs. The rate of TDG formation could thus be higher at higher redshift \citep{Kroupa2010}, which seems to be supported by simulations of gas-rich galaxy mergers \citep{Bournaud2011}.

It it therefore plausible that enough TDGs and GCs form in galaxy-interactions to populate the observed VPOS around the MW.

\subsubsection{Can the tidal scenario explain the different morphologies of the LMC, the dSphs and the GCs?}
If the LMC progenitor galaxy was the origin of the MW satellite system \citep{LyndenBell1976, Pawlowski2011}, this would explain the different morphologies of the Magellanic Clouds compared to the dSph satellites. The LMC would then be the remnant of the initially infalling galaxy, while the majority of MW satellites would have been formed from the tidal debris. Even if the LMC had been born as a TDG, it is expected from numerical simulations \citep{Bournaud2008} that different types of stellar systems are formed: the most massive, rotating TDGs are indistinguishable from dIrr galaxies \citep{Hunter2000}, and systems below $10^8 M_{\sun}$ that \citet{Bournaud2008} describe as super star clusters. These can be interpreted as compact spheroidal galaxies and GCs \citep{Bournaud2008a}. \citet{Bruens2011} have shown that the observed star cluster complexes forming in tidal tails of interacting galaxies can evolve into extended clusters and ultra-compact dwarf galaxies, too. 

All this shows that a large variety of objects can form from tidal debris, such that the different morphologies of satellites observed around the MW do not rule out the tidal scenario of formation of the VPOS.

\subsubsection{Are the varying star formation histories of the MW satellites consistent with a tidal origin?}
The star formation (SF) histories of satellite galaxies born as TDGs can be complex and diverse. TDGs can contain a certain amount of stars from the original galaxy \citep[e.g.][]{Wetzstein2007}, and in addition those stars born when the TDG formed from the tidal-tail material. \citet{Recchi2007} have shown that ongoing SF in TDGs can be expected and gas ejected from the TDG might be re-accreted, leading to another SF period, a process that might be recurring. Additional gas from the tidal tail might be accreted for some time, too. Furthermore, different orbits of the TDGs might influence the SF rate, for example close encounters with a larger galaxy can remove the gas and shut off SF. Therefore, complex SF histories, differing for individual objects born in the same interaction, are expected in TDGs, but no detailed studies on this are available yet. 

With the present limited knowledge about TDG formation and evolution, the SF histories of the MW satellites do not provide definite hints on their origin.

\subsubsection{Can the TDGs be sufficiently long-lived?}
\citet{Bournaud2006} state that only 25 per cent of the TDGs formed in their simulations are long-lived and survive for more than 2 Gyr, but one needs to keep in mind that they only resolve TDGs with masses of at least $10^8 M_{\sun}$. TDGs dropping below this limit are counted as non-surviving, while in nature less massive TDGs are found and the MW satellites typically are less massive, too.

There are observational hints that TDGs are stable. Recently \citet{Duc2011} found three old TDG candidates, aligned along a tidal tail around the galaxy NGC 5557. Their age estimate of these dwarf galaxies (at least 2 Gyr, but maybe up to 5 Gyr) demonstrates that TDGs are long-lived objects which contribute to the population of satellite galaxies. \citet{Galianni2010} discuss that the satellite galaxies of NGC 1097, embedded in a tidal stream, might be old TDGs. 
The stability of TDGs has also been investigated numerically. Using chemodynamical calculations, \citet{Recchi2007} show that dark-matter-free dwarf galaxies easily survive their early star formation phase, which lasts for at least 400 Myr (no longer evolution has been modelled). The long-term stability of dark-matter-free satellites has been demonstrated by \citet{Kroupa1997} and \citet{Klessen1998}. Due to their lack of a massive dark matter halo, TDGs suffer less dynamical friction which should also help them survive when orbiting a more massive host.

TDGs can therefore safely be taken to be long-lived objects.

\subsubsection{How can the large mass-to-light ratios of the satellites be explained?}
\label{sec:masstolight}

TDGs can not contain large amounts of non-baryonic dark matter and should therefore have lower mass-to-light ratios than galaxies embedded in massive dark-matter halos \citep{Barnes1992}.
\citet{Kroupa1997} and \citet{Klessen1998} illustrated that high mass to light ratios of $(M/L)_{\rm{obs}} > 100$\ can be observed in dark-matter free satellite galaxies, even though the intrinsic ratio $(M/L)_{\rm{true}} = 3$\ is low. The reasons are that the galaxies have non-isotropic velocity dispersions, are non-spherical and are not in dynamical equilibrium. This might explain the large observed mass-to-light ratios of some MW satellites, but it is improbable that all of them are out of equilibrium.

Understanding the VPOS in terms of it being a major tidally formed structure is the currently only known viable physical process to naturally account for its phase-space correlation. This then however challenges the validity of the standard cosmological model \citep[see also][]{Kroupa2010, Peebles2010, Famaey2012, Kroupa2012} such that using high $M/L$\ ratios as evidence for the satellite galaxies being dark matter dominated would not be valid. For logical consistency one would need to consider non-Newtonian dynamics.

It is interesting to note in this regard that even the observed rotation curves of young TDGs show signs of missing mass and are surprisingly flat, as was reported by \citet{Bournaud2007}. Using their data, \citet{Gentile2007} have been able to show that the rotation curves are consistent with Modified Newtonian Gravity (MOND). In the MOND-framework, the more massive, ''classical'' dSph satellites naturally obey the Tully-Fisher relation \citep{Tully1977}, as it is a basic consequence of a MONDian gravity law \citep{Milgrom1983, McGaugh2011, Famaey2012}. Indeed, it has already been demonstrated \citep{Brada2000, Angus2008, McGaugh2010} that the $M/L$\ ratios of the MW satellites would be naturally obtained in MOND, with somewhat elevated values for three satellites being the result of tidal effects (\citealt{Kroupa2010}; compare with the Newtonian models by \citealt{Kroupa1997}). \citet{McGaugh2010} have provided intriguing evidence that the unacceptably high mass-to-light ratios of the faint Local Group dwarf galaxies may, in the MONDian framework, result from the dwarfs being out of equilibrium. The preferred alignment of the major axes of ultra-faint dwarfs to the MW centre might hint to a tidal imprint on those satellites \citep{Sand2011}.

Within the standard model of cosmology, current knowledge of TDGs does not support the observed $M/L$\ ratios of the MW satellites, but the issue deserves closer inspection in the future. High observed $M/L$\ ratios therefore do not necessarily exclude a tidal origin, but can be considered the most severe challenge of the tidal scenario \textit{if the dynamics are restricted to the purely Newtonian case}.

\subsubsection{Is it likely that the VPOS has a polar orientation?}
\label{sect:polarlikely}

Assessing the overall likelihood of obtaining a VPOS around a MW-like galaxy demands knowledge of numerous conditions, like those necessary for the formation of TDGs (a gas-rich, probably major, merger) and the probability-distribution of different interaction-geometries. Many of these parameters in turn depend on the underlying dynamics, structure formation and environment. A decisive answer on the overall likelyhood can therefore not be given here, but some parameters are easier to evaluate than others.

First of all, it is to be expected that about 9-12 Gyr ago virtually all galaxies were gas rich. 
If we assume that a galaxy-encounter leads to the formation of a structure of tidal debris, it then comes down to a largely geometrical argument whether an encounter may produce a \textit{polar} structure. For simplicity, it can be assumed that the direction of an infalling galaxy is drawn from an isotropic distribution and that the orientation of tidal debris is mostly defined by the orbital plane of the two interacting galaxies. In that case, every spacial orientation of a disc of tidally formed satellites is equally likely. The probability that the angle between the satellites disc and an arbitrarily oriented Galaxy disc is at least $\chi$\ degree is then given by $\cos \chi$. Thus, the chance to be within $10^\circ$\ to a polar orientation is $\cos 80^\circ = 17.4~\rm{per~cent}$, or more than one in six. In half of the cases, the inclination will be $60^\circ$\ or more. Thus, \textit{if} a structure is formed in a galaxy--galaxy encounter, it is likely to be close to a polar orientation.

In this regard it might be interesting to note that \citet{Bournaud2003} estimate the probability for a galaxy to become a polar ring galaxy due to a merger to be at most 5 per cent. However, they also state that the probability to acquire a polar ring by tidal accretion in a fly-by encounter is significantly larger. This results in their prediction that many polar ring galaxies have not been detected yet as their rings are too faint. One has to exercise care when comparing their probabilities of polar rings to the VPOS. The latter structure is much larger, but also contains less material than a typical polar ring. One might expect that a larger structure is more stable as the matter takes longer to complete one orbit and is more distant from the host galaxy's centre, resulting in a more symmetric potential, but currently this is speculation.

While the tidal scenario can successfully explain the phase-space correlation of the dwarf galaxies, it is currently not possible to give exact numbers stating which alternative for the formation of the MW satellite system is how probable. We do know however that phase-space correlated satellite systems are not uncommon (see Sect. \ref{sect:tdgsindisc}). It is clear that early structure formation and galaxy--galaxy encounters need to be studied in order to assess the overall likelihood of obtaining a VPOS. But as noted in Sec. \ref{sec:masstolight}, if the VPOS can most naturally be understood as a tidal structure, then such work needs to be performed in a non-Newtonian framework for logical consistency. The reason is that if the satellites of the MW would be ancient TDGs, the MW would have no DM-dominated satellites, in contradiction with the predictions of the standard cosmological model.

\subsubsection{Would TDGs follow the mass-size, mass-metallicity and other relations?}
Currently, this question can not be answered satisfactory. No high-resolution studies of the formation and long-time evolution of large numbers of TDGs have been performed. Trends, therefore, can not be investigated in numerical models. Observations do not help either, as the number of known evolved TDGs (in contrast to actively forming ones) is still very low, because evolved TDGs seem to look like ordinary dwarf galaxies.

If MOND (or another form of modified gravity) is valid, its effect on TDGs can be imagined by them being surrounded by ''phantom dark matter''. This possibly results in similar solutions like CDM models. In addition, a stronger gravity would stabilize the TDGs, having the consequence that lifetimes and survival-rates of Newtonian dark-matter free models are only lower limits.

\section{Conclusions}
\label{sect:conclusions}
A vast polar structure (VPOS), consisting of satellite dwarf galaxies, globular clusters classified as young halo (YH) objects, and streams of stars and gas is discovered around the MW.

Applying the disc-fitting algorithm of \citet{Kroupa2010} to three samples of globular clusters around the MW reveals a tight alignment of the normal vector to the disc of YH globular clusters (DoGC$_{\rm{YH}}$) with that of the disc of satellites (DoS). From this it can be concluded that the YH globular clusters, thought to be of extra-galactic origin, are distributed in a similar plane like the MW satellite galaxies. These clusters and the satellite galaxies are therefore distributed in a common VPOS. The plane fitted to them is dominated by the outermost third of YH globular clusters with Galactocentric distances of more than 25 kpc, but fitting the inner 20 clusters independently also gives a distribution closely aligned with the DoS. The probability to find the normal vectors of both the outer and the inner YH cluster distributions this close to the DoS normal is \textit{less than 0.1 per cent} if assuming that the normal vectors are drawn from a uniform distribution on the sphere.
The OH clusters, thought to have formed with the early MW, result in a normal vector not aligned with the DoS. Furthermore, it is much less certain whether their distribution can be fitted with a plane.
As can be expected, GCs classified as Bulge/Disc objects show a normal vector aligning with the MW disc poles, as they are distributed mostly in the plane of the MW.

Using the two-anchor-point method, the normal vectors to streams around the MW centre have been constructed. These are an approximate description of a stream's orbital plane. It is found that half of the streams analysed (7 out of 14) have stream-normal vectors pointing close to the DoS normal vector ($\leq 32^\circ$\ away). The other half have much larger ($\geq 55^\circ$) angular distances from the DoS normal. This hints at a favoured orbital direction of tidal debris, in agreement with them preferentially orbiting within the VPOS. If the stream-normal vectors were drawn from an isotropic distribution, \textit{such a tight alignment with the DoS normal would occur in at most 0.3 per cent of the cases}, ruling out this hypothesis with $3 \sigma$\ confidence. Two out of three streams associated with YH GCs are aligned with the VPOS, showing consistency with the GC disc fitting results, especially as these two clusters are relatively close to the MW and thus do not dominate the found DoGC$_{\rm{YH}}$ normal vector.

The possibility that the clustering of stream-normals is a random structure independent of the VPOS but by chance distributed in a similar direction is thus very unlikely. The same holds true for the distribution of YH GCs in a similar plane like the DoS. In fact, for both the YH GCs and the streams to align with the satellite galaxy DoS has a likelihood of at most 0.1 per cent $\times$\ 0.3 per cent = $3 \times 10^{-4}$\ per cent. Therefore, both findings are significant additions to the DoS, not only in number but also in the type of objects that can be associated with it. Streams originating from dwarf galaxies and globular clusters alike are found close the polar structure, which turns out to be made up not only of satellite galaxies, but also includes globular clusters. This shows for the first time that not only dwarf galaxies are part of and orbit within \citep{Metz2008} this structure, but the same is true for globular clusters. Furthermore, the structure is shown to range deeply into the MW, spreading from the farthest MW satellite galaxies at 250 kpc down to as close as 10 kpc from the Galactic centre for the closest known stream anchor points. Thus, there seems to be a coherent, vast polar structure (VPOS) around the MW which can only have arisen in a major event including the whole Galaxy. 

There are several scenarios trying to explain the VPOS. A tidal origin \citep{Pawlowski2011}, in which the MW satellites have formed in an early, major galaxy-interaction from tidal debris, currently seems to be the most promising possibility. It naturally accounts for the distribution of objects and their (both co- and counter-rotating) orbits in the plane of the interaction, which would have formed dwarf galaxies and star clusters together. In this scenario, the gaseous stream GCN may be an ancient remnant of the tidal tail from which satellite galaxies and star clusters formed. 
From the ages of the YH GCs \citep{Salaris2002, Mackey2004} we deduce this event to have happened 9 to 12 Gyr ago. If the MW satellites are TDGs, we can now not only reconstruct the very first formation events of the MW using OH GCs \citep{Marks2010}, but also the one major encounter that formed the MW bulge, and the satellite galaxies and YH GCs distributed and orbiting in the VPOS. 

The alternative attempts, trying to form a major disc-structure around the MW from cosmological accretion only, suffer from insurmountable problems. Neither group-infall nor filamentary accretion have been shown to be able to account for the very pronounced, polar anisotropy in the spatial distribution of MW satellites in a DoS. In addition, they usually lead to rather diverse orbits which are not as strongly aligned with each other as is in fact observed.
The increased sample-size of objects associated with the VPOS puts even stricter demands on their statistics, such that a pure chance alignment of objects resulting in the observed distribution can be considered ruled out. The implications of this for fundamental physics is discussed in \citet{Kroupa2010}.

\section*{Acknowledgments}

M.S.P. acknowledges support through DFG research grants KR 1635/18-1 and KR 1635/18-2 in the frame of the DFG Priority Programme 1177, \textit{Witnesses of Cosmic History: Formation and evolution of galaxies, black holes, and their environment} and through the Bonn-Cologne Graduate School of Physics and Astronomy. 
We thank Iskren Georgiev for very useful discussions and Peter Weilbacher for the permission to use his 'Dentist's Chair' Galaxy image.

\bibliographystyle{mn2e}
\bibliography{./VPOS}

\label{lastpage}

\end{document}